\documentclass[12pt,a4paper]{article}
\pdfoutput=1
\usepackage{amsmath} 
\usepackage{amsfonts} 
\usepackage{mathtools} 
\usepackage{amssymb} 
\usepackage{mathrsfs}
\usepackage{subfigure} 
\usepackage{graphicx}
\usepackage{color}

\usepackage{jheppub}

\newcommand{\abs}[1]{|#1|} 
\newcommand{\ABS}[1]{\left|#1\right|} 
\newcommand{\re}[1]{\text{Re}\left(#1\right)}
\newcommand{\im}[1]{\text{Im}\left(#1\right)}
\newcommand{\refEQ}[1]{eq.\,\eqref{#1}} 
\newcommand{\refEQS}[1]{eqs.\,\eqref{#1}} 

\newcommand{\Bd}{B^0_d}
\newcommand{\Bdb}{\bar B^0_d}
\newcommand{\BBdmix}{$\Bd$--$\Bdb$}
\newcommand{\BBsmix}{$B^0_s$--$\bar B^0_s$}

\newcommand{\KS}{K_S}
\newcommand{\KL}{K_L}
\newcommand{\KSL}{K_{S,L}}

\newcommand{\KET}[1]{|#1\rangle}
\newcommand{\BRA}[1]{\langle#1|}
\newcommand{\BRAKET}[2]{\langle#1|#2\rangle}
\newcommand{\nof}[1]{B_{\!\nrightarrow#1}}
\newcommand{\noft}[1]{\nof{#1}(t)}
\newcommand{\Knof}[1]{\KET{\nof{#1}}}

\newcommand{\nofp}[1]{B_{\!\nrightarrow#1}^\perp}

\newcommand{\Knofp}[1]{\KET{\nofp{#1}}}

\newcommand{\BKnofp}[1]{\BRAKET{\nofp{#1}}{\nof{#1}}}
\newcommand{\Amp}[1]{A_{#1}}
\newcommand{\AmpB}[1]{\bar{A}_{#1}}
\newcommand{\EAmp}[2]{\mathcal A_{#1}^{#2}}

\newcommand{\MGa}[1]{\langle\Gamma_{#1}\rangle}

\newcommand{\KETo}{|\Bd\rangle}

\newcommand{\KETob}{|\Bdb\rangle}

\newcommand{\KETH}{|B_H\rangle}
\newcommand{\KETL}{|B_L\rangle}

\newcommand{\Lf}{\lambda_{f}}

\newcommand{\Cf}{C_{f}}
\newcommand{\Sf}{S_{f}}
\newcommand{\Rf}{R_{f}}
\newcommand{\Cg}{C_{g}}
\newcommand{\Sg}{S_{g}}
\newcommand{\Rg}{R_{g}}

\newcommand{\C}[1]{C_{#1}}
\renewcommand{\S}[1]{S_{#1}}
\newcommand{\R}[1]{R_{#1}}

\newcommand{\COfont}{\mathscr }
\newcommand{\COC}[3]{\COfont C_{#1}[#2,#3]}
\newcommand{\COS}[3]{\COfont S_{#1}[#2,#3]}
\newcommand{\COCcBASE}{\COfont C_{c}}
\newcommand{\COChBASE}{\COfont C_{h}}
\newcommand{\COScBASE}{\COfont S_{c}}

\newcommand{\COCc}[2]{\COC{c}{#1}{#2}}
\newcommand{\COCh}[2]{\COC{h}{#1}{#2}}
\newcommand{\COSc}[2]{\COS{c}{#1}{#2}}

\newcommand{\FCOCcBASE}{\mathrm C}
\newcommand{\FCOScBASE}{\mathrm S}
\newcommand{\FCOCc}[2]{\FCOCcBASE[#1,#2]}
\newcommand{\FCOSc}[2]{\FCOScBASE[#1,#2]}

\newcommand{\DeltaS}[1]{\Delta S_{\rm #1}}
\newcommand{\DeltaC}[1]{\Delta C_{\rm #1}}
\newcommand{\DST}{\Delta S_{\rm T}^+}
\newcommand{\DSCP}{\Delta S_{\rm CP}^+}
\newcommand{\DSCPT}{\Delta S_{\rm CPT}^+}
\newcommand{\DCT}{\Delta C_{\rm T}^+}
\newcommand{\DCCP}{\Delta C_{\rm CP}^+}
\newcommand{\DCCPT}{\Delta C_{\rm CPT}^+}

\newcommand{\DeltaCh}[1]{\Delta\COfont C_h^{\rm #1}}
\newcommand{\DeltaCc}[1]{\Delta\COfont C_c^{\rm #1}}
\newcommand{\DeltaSc}[1]{\Delta\COfont S_c^{\rm #1}}
\newcommand{\DScT}{\Delta\COfont S_c^{\rm T}}
\newcommand{\DScCP}{\Delta\COfont S_c^{\rm CP}}
\newcommand{\DScCPT}{\Delta\COfont S_c^{\rm CPT}}
\newcommand{\DCcT}{\Delta\COfont C_c^{\rm T}}
\newcommand{\DCcCP}{\Delta\COfont C_c^{\rm CP}}
\newcommand{\DCcCPT}{\Delta\COfont C_c^{\rm CPT}}
\newcommand{\DChT}{\Delta\COfont C_h^{\rm T}}
\newcommand{\DChCP}{\Delta\COfont C_h^{\rm CP}}
\newcommand{\DChCPT}{\Delta\COfont C_h^{\rm CPT}}

\newcommand{\norm}[2]{N_{[#1,#2]}}

\newcommand{\rI}{\hat{\rm I}}
\newcommand{\redIntt}[2]{\rI(#1,#2;t)}
\newcommand{\gBaBar}[3]{\mathbf{g}_{#1,#2}(#3)}
\newcommand{\gBaBart}[2]{\gBaBar{#1}{#2}{t}}

\graphicspath{{Figs/}}

\title{Genuine T, CP, CPT asymmetry parameters for the entangled $\boldsymbol{B_d}$ system}

\author[a]{Jos\'e Bernab\'eu}
\author[a]{Francisco J. Botella}
\author[a]{Miguel Nebot}
\affiliation[a]{Departament de F\'isica Te\`orica and IFIC, Universitat de Val\`encia - CSIC, E-46100, Spain}
\emailAdd{Jose.Bernabeu@uv.es}
\emailAdd{Francisco.J.Botella@uv.es}
\emailAdd{Miguel.Nebot@uv.es}
\preprint{IFIC/16-27}

\vspace{1.0cm}
\abstract{
The precise connection between the theoretical T, CP, CPT asymmetries, in terms of transition probabilities between the filtered neutral meson $B_d$ states, and the experimental asymmetries, in terms of the double decay rate intensities for Flavour-CP eigenstate decay products in a B-factory of entangled states, is established. This allows the identification of genuine Asymmetry Parameters in the time distribution of the asymmetries and their measurability by disentangling genuine and possible fake terms. We express the nine asymmetry parameters -- three different observables for each one of the three symmetries -- in terms of the ingredients of the Weisskopf-Wigner dynamical description of the entangled $B_d$-meson states and we obtain a global fit to their values from the BaBar collaboration experimental results. The possible fake terms are all compatible with zero and the information content of the nine asymmetry parameters is indeed different. The non-vanishing $\DScT =-0.687\pm 0.020 $ and $\DScCP=-0.680\pm 0.021$ are impressive separate direct evidence of Time-Reversal-violation and CP-violation in these transitions and compatible with Standard Model expectations. An intriguing 2$\sigma$ effect for the Re($\theta$) parameter responsible of CPT-violation appears which, interpreted as an upper limit, leads to $|M_{\bar B^0\bar B^0}-M_{B^0B^0}| < 4.0\times 10^{-5}$ eV at 95\% C.L. for the diagonal flavour terms of the mass matrix. It contributes to the CP-violating $\DCcCP$ asymmetry parameter in an unorthodox manner -- in its $\cos(\Delta M\,t)$ time dependence --, and it is accessible in facilities with non-entangled $B_d$'s, like the LHCb experiment.
}
\date{\today}
\begin{document}
\maketitle
 
\section{Introduction\label{SEC:Intro}}
The BaBar collaboration has demonstrated \cite{Lees:2012uka} a 14$\sigma$ direct evidence of Time Reversal violation in the time evolution of the neutral $\Bd$--$\Bdb$\ meson system, independent of CP violation or CPT invariance. This result is independent of any particular dynamical framework for discussing the dynamics of the neutral $\Bd$--$\Bdb$\ system and it is established in terms of asymmetries of observable transition rates. Only the quantum mechanical properties of (i) \emph{entanglement} of the $B_d$ pair before the first decay in a B-Factory, (ii) the decays as \emph{filtering measurements} for the preparation and detection of the initial and final B meson states in the transition, as well as (iii) the time dependence of the double decay rate intensities, are used. The conceptual basis had previously been discussed in refs. \cite{Banuls:1999aj,Banuls:2000ki} and the methodology for an actual experimental analysis given in \cite{Bernabeu:2012ab}, bypassing the need of the T-reversal of the decay. As emphasized by Wolfenstein \cite{Wolfenstein:1999xb,Wolfenstein:1999re}, the T-reverse of a decaying state is not a physical state. 

The transitions of interest are between Flavour and CP eigenstate decay products, with the possibility of having an interference of mixing with no-mixing amplitudes, without any need of absorptive parts, something impossible for transitions between flavour-specific states. In addition, a well defined orthogonality between the meson states filtered by both Flavour and CP eigenstate decay products, the so-called Flavour-Tag and CP-Tag \cite{Banuls:1998mg}, is present under certain conditions \cite{Bernabeu:2012ab,Applebaum:2013wxa,Bernabeu:2013qea}, leading to \emph{genuine} Asymmetries without the contamination of fake terms. The eight different transitions of this kind provide separate independent asymmetries for the T, CP, CPT symmetry transformations. A good time resolution was also a requirement for fixing the precise time ordering of the two decays, as needed for the T and CPT symmetries implemented by antiunitary operators \cite{Wigner:1931,Wigner:1959eng}.

The present study has two main objectives: First, to establish the precise connection between the theoretical asymmetries in terms of transition probabilities for the meson states and the experimental asymmetries in terms of the double decay rate intensities, allowing the identification of genuine Asymmetry Parameters for the model-independent T, CP, CPT time dependent asymmetries. Second, the projection of these Asymmetry Parameters into the ingredients of the Weisskopf-Wigner Approach (WWA) \cite{Weisskopf:1930ps,Lee:1957qq,Enz:1965tr} for the dynamical description of the time evolution of the neutral $\Bd$--$\Bdb$\ system. In particular, we obtain explicit results for:
\begin{enumerate}
\item The construction of the $B_+$, $B_-$ meson states filtered by the CP eigenstate decay products and check their orthogonality.
\item The building of the Asymmetry Parameters for T, CP, CPT in terms of the WWA description. This is of high interest for checking that they are genuine for each of the three symmetries and for demonstrating that the observables for CP and T have, in general, a different information content.
\item The measurability, in the same eight experimental double decay rate intensities, of WWA parameters including possible fake terms in the asymmetries, checking whether the conditions for their absence are met. In general one would be able to disentangle genuine and fake contributions to the asymmetry parameters.
\item The extraction of the values, or limits, of selected WWA parameters, in particular the one responsible of inducing CPT-violation, from a global fit to the observables. Final values for the \emph{genuine} Asymmetry Parameters characterizing the three T, CP, CPT symmetry transformations are given.
\end{enumerate}
The paper is organized as follows: in section \ref{SEC:Evolution} we review some generalities of the \BBdmix\ effective hamiltonian, the time evolution of an initial entangled state and the basic expressions for the double decay rate intensities. In section \ref{SEC:GenuineMR} we discuss in detail under which conditions the Flavour-CP eigenstate decay channels are truly appropriate for time reversal genuine asymmetries. Then, in section \ref{SEC:Babar}, we analyse the experimental asymmetries, normalized as in reference \cite{Lees:2012uka}, focusing on the connection with the time evolution described in section \ref{SEC:Evolution}. In section \ref{SEC:Genuine} we show how the complete genuine asymmetries can be reconstructed beyond ratios through the addition of a single piece of information on the \BBdmix\ mixing. In section \ref{SEC:TrueFake} we address how deviations of the conditions discussed in section \ref{SEC:GenuineMR} can contaminate genuine T and CPT asymmetries, quantifying their effect. In section \ref{SEC:Fit} we present the results of a global fit to the experimental results in terms of the basic WWA parameters introduced in section \ref{SEC:Evolution} and the final results for the genuine asymmetry parameters. Some conclusions are given in section \ref{SEC:Conc}.

\section{Double decay rate time dependent intensities\label{SEC:Evolution}}

The known Weisskopf-Wigner approach (WWA)\cite{Weisskopf:1930ps} for the time evolution of a one level decaying system concluded with the appearance of an absorptive part in the hamiltonian of the Schr\"odinger equation governing its time evolution. The generalization to the two level system gives rise to an effective $2\times 2$ Hamiltonian matrix with an antihermitian part taking care of the decay channels \cite{Lee:1957qq}. These approximations can be obtained using time dependent perturbation theory and have a limited range of validity excluding very short and very long times \cite{Enz:1965tr,Kabir:1968}.

\subsection{The evolution Hamiltonian\label{sSEC:Hamiltonian}}

The effective Hamiltonian of the two meson system \BBdmix\ is $\mathbf{H}=\mathbf{M}-i\mathbf{\Gamma}/2$, where the $2\times 2$ hermitian matrices $\mathbf{M}$ and $\mathbf{\Gamma }$ are respectively the hermitian and the antihermitian parts of $\mathbf{H}$. We follow the notation of \cite{Branco:1999fs} for the eigenvalues
\begin{equation}
\mu_{H,L}=M_{H,L}-\frac{i}{2}\Gamma_{H,L},
\end{equation}
 and eigenvectors
\begin{alignat}{3}
&\mathbf{H}\KETH =\ && \mu _{H}\KETH =\ && p_{H}\KETo+q_{H}\KETob,\label{eq:state:PH}\\
&\mathbf{H}\KETL =\ && \mu _{L}\KETL =\ && p_{L}\KETo-q_{L}\KETob.\label{eq:state:PL}
\end{alignat}
In general, with $\mathbf{H}$ not being a normal operator, $[\mathbf{M},\mathbf{\Gamma}]\neq 0$, the states \eqref{eq:state:PH} and \eqref{eq:state:PL} are not orthogonal. It is convenient to use the averages and differences of masses and widths\footnote{Subindices ``H'' and ``L'' correspond to the ``heavy'' and ``light'' states respectively, and thus $\Delta M>0$ while the sign of $\Delta\Gamma$ is not a matter of convention, it is not fixed here.}:
\begin{alignat}{3}
&\mu &&=\frac{M_{H}+M_{L}}{2}-\frac{i}{2}\frac{\Gamma _{H}+\Gamma _{L}}{2} &&\equiv M-\frac{i}{2}\Gamma, \\
&\Delta \mu &&=M_{H}-M_{L} -\frac{i}{2}\left(\Gamma_{H}-\Gamma_{L}\right) &&\equiv \Delta M-\frac{i}{2}\Delta\Gamma,
\end{alignat}
together with the complex parameters $\theta$ and $q/p$:
\begin{equation}
\frac{q_{H}}{p_{H}}=\frac{q}{p}\sqrt{\frac{1+\theta }{1-\theta}}\,,\qquad
\frac{q_{L}}{p_{L}}=\frac{q}{p}\sqrt{\frac{1-\theta }{1+\theta}}\,,
\end{equation}
and 
\begin{equation}
\delta =\frac{1-|q/p|^2}{1+|q/p|^2}\,.
\end{equation}
It straightforward to check that
\begin{equation}
\label{eq:H:params:01}
\theta =\frac{\mathbf{H}_{22}-\mathbf{H}_{11}}{\Delta \mu}\,,\quad \left(\frac{q}{p}\right)^{2}=\frac{\mathbf{H}_{21}}{\mathbf{H}_{12}}\,.
\end{equation}
Here $\theta$ is a CP and CPT violating complex parameter while $\delta$ violates CP and T. In terms of physical parameters, except for the phase of $q/p$ which is convention dependent\footnote{Although $q/p$ is phase convention dependent, in the CP or T invariant limits, its phase is fixed relative to the convention adopted for the action of the CP operator on $\KET{\Bd}$ and $\KET{\Bdb}$.}, the effective Hamiltonian can be written as \cite{Silva:2000db}
\begin{equation}
\mathbf{H}=
\begin{pmatrix}
\mu -\frac{\Delta \mu }{2}\theta & \frac{p}{q}\frac{\Delta \mu }{2}\sqrt{1-\theta ^{2}} \\ 
\frac{q}{p}\frac{\Delta \mu }{2}\sqrt{1-\theta ^{2}} & \mu +\frac{\Delta \mu}{2}\theta
\end{pmatrix}\,.
\end{equation}
\subsection{The entangled system\label{sSEC:TimeEvol}}
In a B factory operating at the $\Upsilon(4S)$ peak, our initial two-meson state is Einstein-Podolsky-Rosen \cite{Einstein:1935rr} entangled\footnote{See references \cite{Bernabeu:2003ym,Alvarez:2004tj,Alvarez:2006ry} to consider corrections to this assumption.},
\begin{equation}
\label{eq:Entangled:01}
\KET{\Psi_0}=\frac{1}{\sqrt{2}}\Big(\KETo\KETob-\KETob\KETo\Big)=\frac{1}{\sqrt 2(p_Lq_H+p_Hq_L)}\Big(\KETL\KETH-\KETH\KETL\Big),
\end{equation}
which maintains its antisymmetric entangled character in the $\mathbf{H}$ eigenstate basis. This implies the antisymmetric character of the two meson state at all times and for any two independent linear combinations of $\Bd$ and $\Bdb$. The corresponding evolution is therefore given in a simple way. The transition amplitude for the decay of the first state into $\KET{f}$ at time $t_{0}$, and then the second state into $\KET{g}$ at time $t+t_{0}$, is
\begin{equation}
\label{eq:DoubleAmplitude:01}
\BRA{f,t_0;g,t+t_0}T\KET{\Psi_0}=
\frac{e^{-i(\mu_H+\mu_L)t_0}}{\sqrt{2}(p_Lq_H+p_Hq_L)}\Big(e^{-i\mu_H t}\EAmp{f}{L}\EAmp{g}{H}-e^{-i\mu_L t}\EAmp{f}{H}\EAmp{g}{L}\Big),
\end{equation}
where the decay amplitudes of the eigenstates into the final state $f$ are $\EAmp{f}{H,L}\equiv\BRA{f}T\KET{B_{H,L}}$. Squaring and integrating over $t_{0}$, the double decay rate $I(f,g;t)$ is obtained:
\begin{equation}
\label{eq:Intensity:01}
I(f,g;t)=\frac{e^{-\Gamma\, t}}{4\Gamma\abs{p_Lq_H+p_Hq_L}^2}\ABS{e^{i\Delta M\,t/2}e^{\Delta \Gamma\,t/4}\EAmp{f}{H}\EAmp{g}{L}-e^{-i\Delta M\,t/2}e^{-\Delta \Gamma\,t/4}\EAmp{f}{L}\EAmp{g}{H}}^2\,.
\end{equation}
This expression is very useful to realize the following expected symmetry property: up to the global exponential decay factor $e^{-\Gamma\, t}$, the combined transformations $t\rightarrow -t$ and $f\rightleftarrows g$ should be the identity. Expanding the $t$ dependence and taking the approximation $\Delta\Gamma=0$, valid for the neutral $\Bd$ states, one can write:
\begin{equation}
\label{eq:Intensity:02}
I(f,g;t)=e^{-\Gamma\, t}\frac{\MGa{f}\MGa{g}}{\Gamma}\Big\{\COCh{f}{g}+\COCc{f}{g}\cos(\Delta M\,t)+\COSc{f}{g}\sin(\Delta M\,t)\Big\},
\end{equation}
with $\MGa{f}$ defined below. Therefore, starting from an entangled state as in \refEQ{eq:Entangled:01}, and using the Quantum Mechanical evolution in \refEQ{eq:DoubleAmplitude:01}, the following symmetry properties arise from the previous remark:
\begin{equation}
\label{eq:IntensityCoefficients:01}
\COCh{f}{g}=\COCh{g}{f},\quad \COCc{f}{g}=\COCc{g}{f},\quad \COSc{f}{g}=-\COSc{g}{f}. 
\end{equation}
They will play an important role in order to assess the independent observables present in the double decay rate measurements. We define as usual the parameters associated to mixing times decay amplitudes
\begin{equation}
\label{eq:MixingDecay:01}
\lambda_f\equiv\frac{q}{p}\frac{\AmpB{f}}{\Amp{f}},\quad \Cf\equiv \frac{1-\left|\Lf\right|^2}{1+\left|\Lf\right|^2},\quad \Sf\equiv\frac{2\im{\Lf}}{1+\left|\Lf\right|^2},\quad \Rf\equiv\frac{2\re{\Lf}}{1+\left|\Lf\right|^2},
\end{equation}
with $\BRA{f}T\KET{\Bd}\equiv \Amp{f}$, $\BRA{f}T\KET{\Bdb}\equiv \AmpB{f}$ and $\MGa{f}=\frac{1}{2}(\abs{\Amp{f}}^2+\abs{\AmpB{f}}^2)$; note that $\Cf^2+\Sf^2+\Rf^2=1$.\newline 
For flavour specific channels $f=\ell^\pm+X$ ($f=\ell^\pm$ for short in the following), and assuming no wrong lepton charge sign decays, $\C{\ell^\pm}=\pm 1$, $\R{\ell^\pm}=\S{\ell^\pm}=0$, and thus
\begin{align}
\label{eq:IntensityCoefficients:02a}
\COCh{\ell^\pm}{g}&= \norm{\pm}{g}
\left\{
\begin{matrix}
(1+|\theta|^2)(1\mp \Cg)\pm 2\re{\theta^\ast\sqrt{1-\theta^2}}\Rg\\
+|1-\theta^2|(1\pm \Cg)+2\im{\theta^\ast\sqrt{1-\theta^2}}\Sg
\end{matrix}
\right\},\\
\label{eq:IntensityCoefficients:02b}
\COCc{\ell^\pm}{g}&= \norm{\pm}{g}
\left\{
\begin{matrix}
(1-|\theta|^2)(1\mp \Cg)\mp 2\re{\theta^\ast\sqrt{1-\theta^2}}\Rg\\
-|1-\theta^2|(1\pm \Cg)-2\im{\theta^\ast\sqrt{1-\theta^2}}\Sg
\end{matrix}
\right\},\\
\label{eq:IntensityCoefficients:02c}
\COSc{\ell^\pm}{g}&= 2\norm{\pm}{g}\left\{\mp\re{\sqrt{1-\theta^2}}\Sg+\im{\theta}(\pm 1-\Cg)+\im{\sqrt{1-\theta^2}}\Rg\right\},
\end{align}
where $\norm{\pm}{g}=\frac{1\pm\delta}{1-\delta\Cg}$.\\
To close this section, notice that the double decay rate or intensity $I(f,g;t)$ has a trivial normalization by construction: summing over final states $f$ and $g$, and integrating over $t$, we simply have the norm \cite{Roldan:1991,Branco:1999fs} of the initial state $\KET{\Psi_0}$,
\begin{equation}
\int_0^\infty\!\!\!\!\!dt\,\sum_{f,g}\,I(f,g;t)=1\,.
\end{equation}
For later use, it is convenient to introduce the reduced intensity $\rI(f,g;t)$,
\begin{equation}
\label{eq:ReducedIntensity:01}
\rI(f,g;t)\equiv \frac{\Gamma}{\MGa{f}\MGa{g}}I(f,g;t)=e^{-\Gamma\, t}\,\Big\{\COCh{f}{g}+\COCc{f}{g}\cos(\Delta M\,t)+\COSc{f}{g}\sin(\Delta M\,t)\Big\}.
\end{equation}

 
\section{Condition to observe a genuine Motion Reversal asymmetry\label{SEC:GenuineMR}}

The original proposal made in \cite{Banuls:1999aj,Banuls:2000ki} to observe a direct evidence of T violation independently of CP violation at B factories, following reference \cite{Bernabeu:2012ab} and implemented in \cite{Lees:2012uka}, contained three ingredients:
\begin{enumerate}
\item Analyse Time Reversal symmetry in the \BBdmix\ Hilbert space. Therefore, first one defines a reference transition $P_{1}\rightarrow P_{2}(t)$ among meson states and compares with the reversed transition $P_{2}\rightarrow P_{1}(t)$. If the probability that an initially prepared state $P_{1}$, evolved to $P_{1}(t)$, behaves like a $P_{2}$ is 
\begin{equation}
P_{12}(t) =\ABS{\BRA{P_2}U(t,0)\KET{P_1}}^2,
\end{equation}
then the T violating asymmetry proposed was 
\begin{equation}
\label{eq:Asym:Probs:01}
P_{12}(t)-P_{21}(t)\,.
\end{equation}
\item Going beyond the use of $P_{1},P_{2}= \Bd,\Bdb$. If use is made of the transitions $\Bd\rightleftarrows \Bdb$, the corresponding asymmetry is not independent of CP: by construction it is both CP and T violating and very small because it comes from the $\delta$ parameter. They introduced the new reference transition $\Bd\rightarrow B_{+}$ to be compared with $B_{+}\rightarrow \Bd$. In a decay channel with well-defined $CP=+$ where one can neglect CP violation, the reference transition can be measured by looking to decay events $f_{1}$ where a B meson decays to a self-tagging channel of $\Bdb$ and the other B meson decays later to a CP eigenstate $f_{CP=+}$ decay where one can neglect CP violation. The main problem was how to measure the reverse transition.
\item Using the entangled character of the initial state was the crucial ingredient to (i) connect double decay rates with specific meson transitions rates and (ii) to identify the reverse transition. If one assumes that observing a $f_{CP=-}$ one filters in that side a $B_{-}$, then, due to the entanglement, one is tagging the orthogonal state to $B_{-}$ in the opposite side. This state, in the approximation at will, should be a $B_{+}$. In general, from the entangled state \eqref{eq:Entangled:01} we can say that if at time $t_{1}$ we observe in one side the decay product $f$, the (still living) meson at time $t_{1}$ is tagged as the state that does not decay into $f$, $\Knof{f}$,
\begin{equation}
\label{eq:B:nof:01}
\Knof{f}=\frac{1}{\sqrt{\abs{\Amp{f}}^2+\abs{\AmpB{f}}^2}}\left(\AmpB{f}\KETo-\Amp{f}\KETob\right).
\end{equation}
The corresponding orthogonal state $\BKnofp{f}=0$ is given by
\begin{equation}
\label{eq:B:nofp:01}
\Knofp{f}=\frac{1}{\sqrt{\abs{\Amp{f}}^2+\abs{\AmpB{f}}^2}}\left(\Amp{f}^\ast\KETo+\AmpB{f}^\ast\KETob\right),
\end{equation}
and is the one filtered by a decay $f$. What we call the \emph{filtering identity} -- easy to prove \cite{Bernabeu:2013qea,Bernabeu:2015hja} -- defines the precise meaning of the last statement:
\begin{equation}
\ABS{\BRAKET{\nofp{f}}{B_1}}^2=\frac{\ABS{\BRA{f}T\KET{B_1}}^2}{\abs{\Amp{f}}^2+\abs{\AmpB{f}}^2}.
\end{equation}
Note that if $B_{1}=\noft{g}$, the previous quantity is exactly the reduced intensity $\rI(g,f;t)$ introduced in \refEQ{eq:ReducedIntensity:01}:
\begin{equation}
\label{eq:Filtering:ReducedIntensity:01}
\rI(g,f;t)=\frac{\ABS{\BRA{f}T\KET{\noft{g}}}^2}{\abs{\Amp{f}}^2+\abs{\AmpB{f}}^2}=\ABS{\BRAKET{\nofp{f}}{\noft{g}}}^2\,.
\end{equation}
Therefore, here is the precise connection between meson transition probabilities and double decay rates. By measuring $\rI(f_{1},f_{2};t)$, -- from now on we will use the notation $(f_{1},f_{2})$ to refer to the first and the second decays considered -- we are studying probabilities $P_{12}(t)$ for transitions between meson states $(B_{1},B_{2})$ which, as we have seen, are
\begin{equation}
\KET{B_1}=\Knof{f_1}\,,\quad \KET{B_2}=\Knofp{f_2}\,,
\end{equation}
that is transition probabilities for $(B_1,B_2)=(\nof{f_1},\nofp{f_2})$. In order to compare with $P_{21}(t)$, we need to study the reverse transition $(\nofp{f_2},\nof{f_1})$, but the filtering and tagging applied methods do not give us this transition. Two new decay channels $f_{1}^{\prime}$ and $f_{2}^{\prime}$ in the reduced double decay rate $(f_{2}^{\prime},f_{1}^{\prime})$ will give us the transition $(\nof{f_2^\prime},\nofp{f_1^\prime})$; therefore, provided these two new decay channels fulfill the following identity
\begin{equation}
\KET{\nof{f_i^\prime}}=\KET{\nofp{f_i}},
\end{equation}
this new transition $(f_{2}^{\prime},f_{1}^{\prime})$ will give the reversed meson transition. For flavour specific decay channels, assuming no wrong lepton charge sign decays, $\KET{\Bd}=\KET{\nof{\ell^-}}$ and $\KET{\Bdb}=\KET{\nof{\ell^+}}$, this identity is obviously $\KET{\Bdb}=\KET{({\Bd})^\perp}$, and tells us that if $f_{1}=X\ell^{+}\nu_{\ell}$, then $f_{1}^{\prime}=X\ell^{-}\bar\nu_{\ell}$ ($f_{1}=\ell^{+}$ and $f_{1}^{\prime}=\ell^{-}$). The other channel, a CP one, should also satisfy this last equation, which, combined with equations \eqref{eq:B:nof:01} and \eqref{eq:B:nofp:01}, will give the condition these channels should satisfy:
\begin{equation}
\label{eq:Condition:MR:T:00}
\boxed{
\lambda_{f_2}\lambda_{f_2^{\prime}}^{\ast}=-\ABS{\frac{q}{p}}^{2}
}\,.
\end{equation}
The originally proposed decay channels  $f_2=J/\psi K_+$ and $f_2^{\prime}=J/\psi K_-$ satisfy this condition,
 where $K_\pm$ are the neutral kaon states filtered by the CP eigenstate decay channels. Consequently, the states $B_\mp$ are well defined and given by equation \eqref{eq:B:nofp:01} for each of the two decay channels. 
From now on, we use $\KS$ for $K_+$ and $\KL$ for $K_-$ since it is an accurate approximation up to CP violation in the kaon system. 
Taking into account that $\lambda_{J/\psi \KS}\equiv \lambda_{\KS}\sim \ABS{\frac{q}{p}}e^{-i2\beta}$ and $\lambda_{J/\psi \KL}\equiv\lambda_{\KL}\sim -\ABS{\frac{q}{p}} e^{-i2\beta}$,
 to control potential deviations from condition \eqref{eq:Condition:MR:T:00}, we will use the general parameterisation
\begin{equation}
\label{eq:Lambdas:Param:01}
\lambda_{\KS}=\ABS{\frac{q}{p}}\,\rho\,(1+\epsilon_\rho)\,e^{-i(2\beta+\epsilon_\beta)}\,,\quad 
\lambda_{\KL}=-\ABS{\frac{q}{p}}\,\frac{1}{\rho}\,(1+\epsilon_\rho)\,e^{-i(2\beta-\epsilon_\beta)}\,,
\end{equation}
in terms of the real parameters $\{\rho,\beta,\epsilon_\rho,\epsilon_\beta\}$. Therefore, by properly comparing double decay rates corresponding to two channels, one from $\{\ell^{+},\ell^{-}\}$ and the other from $\{J/\psi \KS,J/\psi \KL\}$ ($\KS$ and $\KL$ for short in the following), we will be able to measure genuine time-reverse processes provided
\begin{equation}
\label{eq:Condition:MR:T:Params:01}
\epsilon_\rho=0\,,\quad \epsilon_\beta=0\,.
\end{equation}
Any deviation of this measurable relation produces some contamination in time reversal asymmetries and therefore should be conveniently subtracted out. It is important to notice that \refEQ{eq:Condition:MR:T:00} is fulfilled even if $\rho\neq 1$. It has to be pointed out that \refEQ{eq:Condition:MR:T:00} guarantees that the considered channels allow to truly compare the transition $P_{2}\rightarrow P_{1}(t)$ with the reversed transition $P_{1}\rightarrow P_{2}(t)$. Nevertheless, in order to ensure that this motion reversal asymmetry is truly a time reversal asymmetry, one has to use decay channels $f$ such that in the limit of T invariance, $\S{f}=0$ \cite{Bernabeu:2013qea,Bernabeu:2015hja}. For CP eigenstates, T invariance implies $\S{f}=0$ provided there is no CPT violation in the corresponding decay amplitude, in accordance with the analysis in reference \cite{Applebaum:2013wxa}. This is equivalent to no CP violation in the decay, in the T invariant limit, giving, in addition to \refEQ{eq:Condition:MR:T:Params:01}, the condition $\rho=1$. We therefore conclude that we should perform the data analysis with arbitrary parameters $\rho$, $\epsilon_{\rho}$ and $\epsilon_{\beta}$ and that any deviation from 
\begin{equation}
\label{eq:T:condition:params}
\boxed{\rho=1\,,\ \epsilon_\rho=0\,,\ \epsilon_\beta=0}\,,
\end{equation}
will be a source of fake T violation that should be subtracted out. Notice that in the absence of CP violation in the decays that filter the states $B_{\pm}$, these states would be orthogonal, implying \refEQ{eq:T:condition:params}, and therefore the orthogonality condition in equation \eqref{eq:Condition:MR:T:00} would be automatically satisfied. Before ending this section it is convenient to clarify that in the absence of wrong flavour decays in $\Bd\to J/\psi K^{0}$ and $\Bdb\to J/\psi \bar{K}^{0}$, one has $\lambda_{\KS}+\lambda_{\KL}=0$ (see \cite{Grossman:2002bu}); in our parameterisation, this implies
\begin{equation}\label{eq:NWS:02}
\rho =1\,,\ \epsilon_{\beta}=0\,,
\end{equation}
clearly showing full compatibility among the condition in \refEQ{eq:Condition:MR:T:00} and the absence of wrong flavour decays. Using more conventional notation in terms of $\C{\KS}$, $\C{\KL}$, $\S{\KS}$, $\S{\KL}$, $\R{\KS}$ and $\R{\KL}$ (\refEQ{eq:MixingDecay:01}), no wrong flavour decays imply 
\begin{equation}
\label{eq:T:condition:CSR:01}
\C{\KS}-\C{\KL}=0\,,\ \S{\KS}+\S{\KL}=0\,,\ \R{\KS}+\R{\KL}=0\,.
\end{equation}
If we impose in addition \refEQ{eq:Condition:MR:T:00}, we also have
\begin{equation}
\label{eq:T:condition:C:02}
C_{\KS}=C_{\KL}=\delta\,.
\end{equation}
\end{enumerate}

\section{The BaBar normalization and the independent asymmetries\label{SEC:Babar}}

To avoid strong dependences on the detection efficiencies in the different channels, reference \cite{Lees:2012uka}, instead of measuring $\COCh{f}{g}$, $\COCc{f}{g}$ and $\COSc{f}{g}$ in \refEQ{eq:Intensity:02} or \refEQ{eq:ReducedIntensity:01},
fixed the normalization of the constant term and used the normalized decay intensity
\begin{equation}
\mathbf{g}_{f,g}(t)\propto e^{-\Gamma\, t}\left\{1+\FCOCc{f}{g}\cos(\Delta M\,t)+\FCOSc{f}{g}\sin(\Delta M\,t)\right\},
\end{equation}
in such a way that two quantities, 
\begin{equation}
\label{eq:IntensityCoefficients:03}
\FCOCc{f}{g}=\frac{\COCc{f}{g}}{\COCh{f}{g}},\quad \FCOSc{f}{g}=\frac{\COSc{f}{g}}{\COCh{f}{g}},
\end{equation}
are measured for each pair $(f,g)$. Following \refEQ{eq:IntensityCoefficients:01}, they verify
\begin{equation}
\label{eq:IntensityCoefficients:04}
\FCOCc{f}{g}=\FCOCc{g}{f},\quad \FCOSc{f}{g}=-\FCOSc{g}{f}\,.
\end{equation}
We are interested in the study of the genuine discrete asymmetries that can be constructed combining one flavour specific channel and one CP channel. Starting from one reference transition, we can generate another three by means of T, CP and CPT transformations. 
%
%
It turns out that because of the relation \eqref{eq:IntensityCoefficients:04}, these four transitions $\Bdb\to B_{-}$, $B_{-}\to\Bdb$, $\Bd\to B_{-}$ and $B_{-}\to \Bd$ saturate all the independent parameters that can be measured with one flavour specific and one CP decays. In Table \ref{TAB:Transitions} we present the meson state transitions and the corresponding decay channels, and we see how with one reference transition and its discrete symmetry transformed ones, all the independent parameters are saturated: the order below the $\mathbf{g}_{f,g}(t)$ column makes clear that the parameters of these transitions are related to the ones appearing in the column $\mathbf{g}_{g,f}(t)$. We conclude that only eight parameters are independent: they are the $\FCOCc{f}{g}$ and $\FCOSc{f}{g}$ corresponding to the decays $(\ell^{+},\KS)$, $(\KL,\ell^{-})$, $(\ell^{-},\KS)$ and $(\KL,\ell^{+})$. Of course, there are at least two independent ways of measuring the same parameter by means of the time-ordering of the two decays. This operation is not a symmetry transformation from the left to the right-hand side of Table \ref{TAB:Transitions}; in order to interpret the information it is very important to know exactly the number of independent parameters in a general framework.
\begin{table}[h!tb]
\begin{center}
\begin{tabular}{|c|c|c||c|c|c|}
\hline
\multicolumn{2}{|r|}{Transition} & $\mathbf{g}_{f,g}(t)$ & $\mathbf{g}_{g,f}(t)$ & \multicolumn{2}{|l|}{Transition} \\ \hline
Reference & $\Bdb\to B_-$ & $(\ell^+,\KS)$ & $(\KS,\ell^+)$ & $B_+\to\Bd$ & Reference\\ \hline
T-transformed & $B_-\to\Bdb$ & $(\KL,\ell^-)$ & $(\ell^-,\KL)$ & $\Bd\to B_-$ & T-transformed\\ \hline
CP-transformed & $\Bd\to B_-$ & $(\ell^-,\KS)$ & $(\KS,\ell^-)$ & $B_+\to\Bdb$ & CP-transformed\\ \hline
CPT-transformed & $B_-\to\Bd$ & $(\KL,\ell^+)$ & $(\ell^+,\KL)$ & $\Bdb\to B_+$ & CPT-transformed\\ \hline
\end{tabular}
\caption{Double decay channels, the associated filtered meson states and their transformed transitions under the three discrete symmetries.\label{TAB:Transitions}}
\end{center}
\end{table}

\noindent The authors in reference \cite{Bernabeu:2012ab} proposed the construction of several CP, T or CPT asymmetries as BaBar did, in order to present genuine and model independent tests of these symmetries. By now, it should be clear that only \emph{six} independent asymmetries can be constructed out of the eight independent parameters. The three time dependent asymmetries are
\begin{align}
A_{\rm T}(t)&= \gBaBart{\KL}{\ell^-}-\gBaBart{\ell^+}{\KS}\,,\label{eq:AsymBabar:Int:01a}\\
A_{\rm CP}(t) &= \gBaBart{\ell^-}{\KS}-\gBaBart{\ell^+}{\KS}\,,\label{eq:AsymBabar:Int:01b}\\
A_{\rm CPT}(t) &= \gBaBart{\KL}{\ell^+}-\gBaBart{\ell^+}{\KS}\,,\label{eq:AsymBabar:Int:01c}
\end{align}
which can be explicitely expanded as
\begin{equation}
\label{eq:AsymBabar:Int:02}
A_{\rm S}(t)=e^{-\Gamma t}
\left\{ \DeltaC{S}[\ell^{+},\KS]\,\cos(\Delta M\,t) +\DeltaS{S}[\ell^{+},\KS]\,\sin(\Delta M\,t)\right\},\, {\rm S}={\rm T}, {\rm CP}, {\rm CPT}\,,
\end{equation}
where
\begin{align}
\DCT\equiv \DeltaC{T}[\ell^{+},\KS] = \FCOCc{\KL}{\ell^-}-\FCOCc{\ell^+}{\KS}\,,\label{eq:AsymBabar:C:01a}\\
\DCCP\equiv \DeltaC{CP}[\ell^{+},\KS] = \FCOCc{\ell^-}{\KS}-\FCOCc{\ell^+}{\KS}\,,\label{eq:AsymBabar:C:01b}\\
\DCCPT\equiv \DeltaC{CPT}[\ell^{+},\KS] = \FCOCc{\KL}{\ell^+}-\FCOCc{\ell^+}{\KS}\,,\label{eq:AsymBabar:C:01c}
\end{align}
\begin{align}
\DST\equiv \DeltaS{T}[\ell^{+},\KS] = \FCOSc{\KL}{\ell^-}-\FCOSc{\ell^+}{\KS}\,,\label{eq:AsymBabar:S:01a}\\
\DSCP\equiv \DeltaS{CP}[\ell^{+},\KS] = \FCOSc{\ell^-}{\KS}-\FCOSc{\ell^+}{\KS}\,,\label{eq:AsymBabar:S:01b}\\
\DSCPT\equiv \DeltaS{CPT}[\ell^{+},\KS] = \FCOSc{\KL}{\ell^+}-\FCOSc{\ell^+}{\KS}\,,\label{eq:AsymBabar:S:01c}
\end{align}
are the six independent asymmetries that can be constructed (we use the same notation of reference \cite{Lees:2012uka} for easy comparison). To appreciate the difference among asymmetries that in a CPT invariant world would be equivalent, we can write them expanding to linear order in $\re{\theta}$, $\im{\theta}$:
\begin{align}
\DST &\simeq \S{\KS}-\S{\KL}-\re{\theta}(\S{\KS}\R{\KS}+\S{\KL}\R{\KL})\nonumber\\ &\hspace{4cm}+\im{\theta}(\S{\KS}^2-\S{\KL}^2+\C{\KS}+\C{\KL}),\label{eq:aproxAsymmetries:ST}\\
\DSCP &\simeq 2\S{\KS}+2\im{\theta}(\S{\KS}^2-1),\label{eq:aproxAsymmetries:SCP}\\
\DSCPT &\simeq \S{\KL}+\S{\KS}-\re{\theta}(\S{\KL}\R{\KL}+\S{\KS}\R{\KS})\nonumber\\ &\hspace{4cm}+\im{\theta}(-2+\S{\KS}^2+\S{\KL}^2+\C{\KS}+\C{\KL}),\label{eq:aproxAsymmetries:SCPT}
\end{align}
\begin{align}
\DCT &\simeq \C{\KS}+\C{\KL}+\re{\theta}(\R{\KS}(1-\C{\KS})+\R{\KL}(1+\C{\KL}))\nonumber\\ &\hspace{4cm} +\im{\theta}(\S{\KL}(1+\C{\KL})-\S{\KS}(1-\C{\KS})),\label{eq:aproxAsymmetries:CT}\\
\DCCP &\simeq 2\C{\KS}+2\re{\theta}\R{\KS}+2\im{\theta}\S{\KS}\C{\KS},\label{eq:aproxAsymmetries:CCP}\\
\DCCPT &\simeq \C{\KS}-\C{\KL}+\re{\theta}(\R{\KS}(1-\C{\KS})-\R{\KL}(1-\C{\KL}))\nonumber\\ &\hspace{4cm}+\im{\theta}(\S{\KL}(1-\C{\KL})-\S{\KS}(1-\C{\KS})).\label{eq:aproxAsymmetries:CCPT}
\end{align}
No matter whether CPT Violation is expected to be small,
 conceptually it is very important to emphasize that $\DST\neq\DSCP$ for several reasons. We have seen that for $\DST$ to be a true T violating asymmetry, \refEQS{eq:T:condition:CSR:01} and \eqref{eq:T:condition:C:02} should be fulfilled. Therefore, the dominant term in equations \eqref{eq:aproxAsymmetries:ST} and \eqref{eq:aproxAsymmetries:SCP} should be equal: $\S{\KS}-\S{\KL}=2\S{\KS}$. But in general $\DST$ and $\DSCP$ differ by terms that are CPT violating and CP invariant in $\DST$, and by terms that are CPT violating and T invariant in $\DSCP$. Only the pieces that do not depend on $\theta$ are identical. Similarly for $\DCT\neq \DCCP$: in order for $\DCT$ to be a true T violating asymmetry, we need $\C{\KS}+\C{\KL}=2\C{\KS}=2\delta$, and thus $\DCT$ and $\DCCP$ are again equal up to CPT violation in the mixing: they differ by terms that are CPT violating and CP invariant in $\DCT$ and by terms that are CPT violating and T invariant in $\DCCP$.
It is a very important check to realize that both $\DSCPT$ and $\DCCPT$ only contain pieces proportional to CPT violating parameters: they have terms proportional to the $\theta$ parameter controlling the amount of CPT violation in the mixing. $\DSCPT$ also contains $\S{\KS}+\S{\KL}$, which should be equal to zero if $\DSCPT$ is a true CPT asymmetry (the same condition in \refEQ{eq:Condition:MR:T:00} for a true T asymmetry). Finally, $\DCCPT$ contains $\C{\KS}-\C{\KL}$, which should vanish provided \refEQ{eq:Condition:MR:T:00} is fulfilled and there is no CPT violation in the decay.

\section{Genuine asymmetry parameters\label{SEC:Genuine}}
The time-dependent reduced intensity $\rI(f,g;t)$ involves three coefficients $\COChBASE$, $\COCcBASE$ and $\COScBASE$ in \refEQS{eq:IntensityCoefficients:02a}, \eqref{eq:IntensityCoefficients:02b} and \eqref{eq:IntensityCoefficients:02c}; nevertheless, as mentioned before, the analysis of reference \cite{Lees:2012uka} focused on the ratios $\FCOCcBASE=\COCcBASE/\COChBASE$ and $\FCOScBASE=\COScBASE/\COChBASE$ in \refEQ{eq:IntensityCoefficients:03}. Although from the experimental point of view those ratios might be more appropriate, from the theoretical point of view, access to the three independent coefficients would be more desirable: for instance, while an asymmetry in the ratios does imply a symmetry violation, no asymmetry in the ratios may nevertheless come from asymmetries in both the numerator and the denominator. Obtaining the three independent coefficients $\COChBASE$, $\COCcBASE$ and $\COScBASE$ for each pair of decay channels might be particularly interesting for asymmetries in the ratios with values that are, within uncertainties, compatible with zero, like e.g. CPT asymmetries. Is that programme possible? Fortunately, using input information for $\abs{q/p}$ or, equivalently $\delta$, it can be achieved.
First, from \refEQS{eq:IntensityCoefficients:02a} and \eqref{eq:IntensityCoefficients:02b}, we have
\begin{equation}
\label{eq:ChCcSUM:01}
\COCh{\ell^\pm}{\KSL}+\COCc{\ell^\pm}{\KSL}=\frac{(1\pm\delta)(1\mp\C{\KSL})}{2(1-\delta\C{\KSL})}=\COCh{\ell^\pm}{\KSL}\left(1+\FCOCc{\ell^\pm}{\KSL}\right)\,.
\end{equation}
Equation \eqref{eq:ChCcSUM:01} is interpreted in the following way: while $\FCOCc{\ell^\pm}{\KSL}$ and $\C{\KSL}$ will be constrained or extracted from the data, through the addition of $\delta$, we can also compute $\COCh{\ell^\pm}{\KSL}$, and thus $\COCc{\ell^\pm}{\KSL}$ and $\COSc{\ell^\pm}{\KSL}$ separately. It is then possible to build T, CP and CPT complete time-dependent asymmetries analog to \refEQS{eq:AsymBabar:Int:01a}, \eqref{eq:AsymBabar:Int:01b} and \eqref{eq:AsymBabar:Int:01c},
\begin{align}
\mathscr A_{\rm T}(t)&= \redIntt{\KL}{\ell^-}-\redIntt{\ell^+}{\KS}\,,\label{eq:AsymGenuine:Int:01a}\\
\mathscr A_{\rm CP}(t) &= \redIntt{\ell^-}{\KS}-\redIntt{\ell^+}{\KS}\,,\label{eq:AsymGenuine:Int:01b}\\
\mathscr A_{\rm CPT}(t) &= \redIntt{\KL}{\ell^+}-\redIntt{\ell^+}{\KS}\,,\label{eq:AsymGenuine:Int:01c}
\end{align}
which can also be expanded as
\begin{equation}
\label{eq:AsymGenuine:Int:02}
\mathscr A_{\rm S}(t)=e^{-\Gamma t}
\left\{ \DeltaCh{S}+\DeltaCc{S}\,\cos(\Delta M\,t) +\DeltaSc{S}\,\sin(\Delta M\,t)\right\}\,,\qquad {\rm S}={\rm T}, {\rm CP}, {\rm CPT}\,.
\end{equation}
We refer to $\DeltaCh{S}$, $\DeltaCc{S}$ and $\DeltaSc{S}$ in these asymmetries as ``genuine asymmetry parameters'' since they are the ones which collect the full time-dependent difference of probabilities in transitions among meson states given in \refEQ{eq:Asym:Probs:01}. 
For completeness, we write down the general expressions in \refEQS{eq:IntensityCoefficients:02a}, \eqref{eq:IntensityCoefficients:02b} and \eqref{eq:IntensityCoefficients:02c} expanded up to linear order in $\theta$ and $\delta$,
\begin{align}
\label{eq:Genuine:Ch:01}
\COCh{\ell^\pm}{g}&=\frac{1}{2}\big\{1+\delta(\Cg\pm 1)\pm\re{\theta}\Rg-\im{\theta}\Sg\big\},\\
\label{eq:Genuine:Cc:01}
\COCc{\ell^\pm}{g}&=\frac{1}{2}\big\{\mp\Cg+\delta\Cg(\mp\Cg-1)\mp\re{\theta}\Rg+\im{\theta}\Sg\big\},\\
\label{eq:Genuine:Sc:01}
\COSc{\ell^\pm}{g}&=\frac{1}{2}\big\{\mp\Sg+\delta\Sg(\mp\Cg-1)+\im{\theta}(\pm 1-\Cg)\big\},
\end{align}
from which the genuine asymmetry parameters in the coefficients $\COChBASE$, $\COCcBASE$ and $\COScBASE$, up to linear order in $\theta$ and $\delta$, follow:
\begin{alignat}{2}
\DChT&\equiv\COCh{\KL}{\ell^-}-\COCh{\ell^+}{\KS}=\nonumber\\ &\frac{1}{2}\big\{\delta(\C{\KL}-\C{\KS}-2)
-\re{\theta}(\R{\KL}+\R{\KS})+\im{\theta}(\S{\KS}-\S{\KL})\big\},\label{eq:GenuineAsym:Ch:T:01}\\
\DChCP&\equiv\COCh{\ell^-}{\KS}-\COCh{\ell^+}{\KS}=-\big\{\delta+\re{\theta}\R{\KS}\big\},\label{eq:GenuineAsym:Ch:CP:01}\\
\DChCPT&\equiv\COCh{\KL}{\ell^+}-\COCh{\ell^+}{\KS}=\nonumber\\ &\frac{1}{2}\big\{\delta(\C{\KL}-\C{\KS})
+\re{\theta}(\R{\KL}-\R{\KS})+\im{\theta}(\S{\KS}-\S{\KL})\big\},\label{eq:GenuineAsym:Ch:CPT:01}
\end{alignat}
\begin{alignat}{2}
\DCcT&\equiv\COCc{\KL}{\ell^-}-\COCc{\ell^+}{\KS}=\frac{1}{2}\big\{\C{\KS}+\C{\KL}\nonumber\\ 
+\delta (&\C{\KL}(\C{\KL}-1)+\C{\KS}(\C{\KS}+1)) +\re{\theta}(\R{\KS}+\R{\KL})+\im{\theta}(\S{\KL}-\S{\KS})\big\},\label{eq:GenuineAsym:Cc:T:01}\\
\DCcCP&\equiv\COCc{\ell^-}{\KS}-\COCc{\ell^+}{\KS}=\big\{\C{\KS}+\delta\C{\KS}^2+\re{\theta}\R{\KS}\big\},\label{eq:GenuineAsym:Cc:CP:01}\\
\DCcCPT&\equiv\COCc{\KL}{\ell^+}-\COCc{\ell^+}{\KS}=\frac{1}{2}\big\{\C{\KS}-\C{\KL}\nonumber\\ 
+\delta (&\C{\KS}(\C{\KS}+1)-\C{\KL}(\C{\KL}+1))+\re{\theta}(\R{\KS}-\R{\KL})+\im{\theta}(\S{\KL}-\S{\KS})\big\},\label{eq:GenuineAsym:Cc:CPT:01}
\end{alignat}
\begin{alignat}{2}
\DScT&\equiv\COSc{\KL}{\ell^-}-\COSc{\ell^+}{\KS}=\frac{1}{2}\big\{\S{\KS}-\S{\KL}\nonumber\\ +\delta(&\S{\KS}(1+\C{\KS})+\S{\KL}(1-\C{\KL}))
+\im{\theta}(\C{\KS}+\C{\KL})\big\},\label{eq:GenuineAsym:Sc:T:01}\\
\DScCP&\equiv\COSc{\ell^-}{\KS}-\COSc{\ell^+}{\KS}=\big\{\S{\KS}+\delta\S{\KS}\C{\KS}-\im{\theta}\big\},\label{eq:GenuineAsym:Sc:CP:01}\\
\DScCPT&\equiv\COSc{\KL}{\ell^+}-\COSc{\ell^+}{\KS}=\frac{1}{2}\big\{\S{\KS}+\S{\KL}\nonumber\\ +\delta(&\S{\KS}(\C{\KS}+1)+\S{\KL}(\C{\KL}+1))
-\im{\theta}(2+\C{\KS}-\C{\KL})\big\}.\label{eq:GenuineAsym:Sc:CPT:01}
\end{alignat}
It is important to stress from \eqref{eq:ChCcSUM:01} that it has a straightforward physical interpretation: from \refEQ{eq:ReducedIntensity:01}, the reduced intensity prior to any time evolution is
\begin{equation}
\rI(f,g;0)=\COCh{f}{g}+\COCc{f}{g}\,.
\end{equation}
Following the filtering identity in \refEQ{eq:Filtering:ReducedIntensity:01}, this is simply the overlap between $\KET{\nofp{g}}$ and $\KET{\nof{f}}$:
\begin{equation}
\label{eq:tzero:01}
\rI(f,g;0)=\ABS{\BRAKET{\nofp{g}}{\nof{f}}}^2=\frac{\ABS{\AmpB{f}\Amp{g}-\Amp{f}\AmpB{g}}^2}{(\abs{\Amp{f}}^2+\abs{\AmpB{f}}^2)(\abs{\Amp{g}}^2+\abs{\AmpB{g}}^2)}\,.
\end{equation}
Furthermore, it can be easily seen that, if the condition in \refEQ{eq:Condition:MR:T:00} for a genuine Motion Reversal measurement is verified, $\mathscr A_{\rm T}(0)=0$, since
\begin{multline}
\mathscr A_{\rm T}(0)=\rI({\KL},{\ell^-};0)-\rI({\ell^+},{\KS};0)=\ABS{\BRAKET{\nofp{\KL}}{\nof{\ell^-}}}^2-\ABS{\BRAKET{\nofp{\ell^+}}{\nof{\KS}}}^2\\ =\frac{\abs{\Amp{\KL}}^2}{\abs{\Amp{\KL}}^2+\abs{\AmpB{\KL}}^2}-\frac{\abs{\AmpB{\KS}}^2}{\abs{\Amp{\KS}}^2+\abs{\AmpB{\KS}}^2}\,,
\end{multline}
and thus
\begin{equation}
\mathscr A_{\rm T}(0)=0\Leftrightarrow \frac{\abs{\AmpB{\KL}}^2}{\abs{\Amp{\KL}}^2}=\frac{\abs{\Amp{\KS}}^2}{\abs{\AmpB{\KS}}^2}\,,\ \text{while}\ \lambda_{\KL}\lambda_{\KS}^\ast=-\ABS{\frac{q}{p}}^2\Leftrightarrow \frac{\AmpB{\KL}}{\Amp{\KL}}=-\frac{\Amp{\KS}^\ast}{\AmpB{\KS}^\ast}\,.
\end{equation}
This is consistent with the intuitive requirement that a genuine Motion Reversal asymmetry cannot be already present at $t=0$, i.e. in the absence of time evolution. Concerning CPT, $\mathscr A_{\rm CPT}(0)=0$ on the same grounds that $\mathscr A_{\rm T}(0)=0$, once the CP properties of the decay states and the absence of CP violation in the decays are considered.
One final comment is in order: attending to the previous results, the presence of $\delta$ in \refEQ{eq:ChCcSUM:01}, that is at $t=0$, is a priori surprising, since it is solely related to the \BBdmix\ mixing; this is simply an artifact due to the use of the mixing times decay quantities in \refEQ{eq:MixingDecay:01}, as illustrated by the absence of $\delta$ in \refEQ{eq:tzero:01}. In any case we should keep the normalization of \refEQ{eq:ChCcSUM:01} since we also want to measure deviations from \refEQ{eq:T:condition:params}.

\section{Genuine T-reverse and fake asymmetries\label{SEC:TrueFake}}

In section \ref{SEC:GenuineMR} we have discussed how asymmetries like \refEQS{eq:AsymBabar:Int:01a} and \eqref{eq:AsymBabar:Int:01c} are ``contaminated'', i.e. can receive contributions which are not truly T-violating; this also applies to the genuine asymmetry parameters introduced in section \ref{SEC:Genuine}. It occurs when the conditions in \refEQ{eq:T:condition:params} are not fulfilled. The question is, how can we disentangle \emph{fake} effects in T and CPT asymmetries due to deviations from the requirements of \refEQ{eq:T:condition:params}? We illustrate the reasoning using, for example, the asymmetry $\DScT$ in \refEQ{eq:GenuineAsym:Sc:T:01}. First, we remind the reader that in terms of all parameters involved in the problem -- $\delta$, $\rho$, $\beta$, $\epsilon_\rho$ and $\epsilon_\beta$ in \refEQ{eq:Lambdas:Param:01}, plus the complex $\theta$ parameter --, $\DScT$ is simply a function $\DScT(\rho,\beta,\epsilon_\rho,\epsilon_\beta,\delta,\theta)$. $\DScT$ would be a \emph{true} T-violation asymmetry if $\epsilon_\rho=\epsilon_\beta=0$ and $\rho=1$ (\refEQ{eq:T:condition:params}). It is then possible to do the following separation, at each point in parameter space, when performing a fit to the observables:
\begin{multline}
\label{eq:TrueFake:01}
\DScT(\rho,\beta,\epsilon_\rho,\epsilon_\beta,\delta,\theta)=\Big[\DScT(\rho,\beta,\epsilon_\rho,\epsilon_\beta,\delta,\theta)-\DScT(1,\beta,0,0,\delta,\theta)\Big]\\ +\DScT(1,\beta,0,0,\delta,\theta)\,.
\end{multline}
The term within square brackets,
\begin{equation}
\label{eq:TrueFake:02}
\DScT(\rho,\beta,\epsilon_\rho,\epsilon_\beta,\delta,\theta)-\DScT(1,\beta,0,0,\delta,\theta)\,,
\end{equation}
has exactly the desired properties for the \emph{fake} contribution: independently of $\beta$, $\delta$ and $\theta$, it vanishes when the conditions \refEQS{eq:Condition:MR:T:00} and \eqref{eq:T:condition:params} are fulfilled. Then, the last term,
\begin{equation}
\label{eq:TrueFake:02}
\DScT(1,\beta,0,0,\delta,\theta)\,,
\end{equation}
is the truly T-violating contribution, the \emph{genuine T-reverse} one. It is then possible to quantify the amounts of fake and genuine T-reverse contributions to T and CPT asymmetries like $\DST$, $\DCT$, $\DSCPT$, $\DCCPT$, and also, of course, to the T and CPT genuine asymmetry parameters involving the individual $\COChBASE$, $\COCcBASE$ and $\COScBASE$ coefficients. They are explicitely shown in the results of the fit in section \ref{SEC:Fit}.
In terms of the parameters $\delta$, $\rho$, $\beta$, $\epsilon_\rho$ and $\epsilon_\beta$ in \refEQ{eq:Lambdas:Param:01}, the genuine T-reverse asymmetries are simply obtained for
\begin{equation}
\label{eq:GenuineAsy:TruePars:01}
\left\{\begin{matrix}\C{\KS}\\ \C{\KL}\end{matrix}\right\}\,\to\delta,\quad
\left\{\begin{matrix}\S{\KS}\\ -\S{\KL}\end{matrix}\right\}\,\to -\sqrt{1-\delta^2}\sin 2\beta,\quad 
\left\{\begin{matrix}\R{\KS}\\ -\R{\KL}\end{matrix}\right\}\,\to\sqrt{1-\delta^2}\cos 2\beta\,.
\end{equation}

\section{Results\label{SEC:Fit}}

Following the ideas developed in the previous sections, we now present results obtained from a global fit to the available experimental information. First, we discuss in section \ref{sSEC:Fit:Gen} the basics of the global fit and the main results, including in particular the new best determination of the real part of the CPT violating parameter $\theta$. Then, in section \ref{sSEC:Fit:Various}, we illustrate and discuss several specific aspects of the results: the difference between CP and T asymmetries -- as discussed in section \ref{SEC:Babar} --, the separation of genuine T-reverse asymmetries and fake contributions, and finally the sensitivity of different asymmetries to $\re{\theta}$ and $\im{\theta}$.

\subsection{Global fit\label{sSEC:Fit:Gen}}
\begin{figure}[h]
\begin{center}
\includegraphics[width=0.4\textwidth]{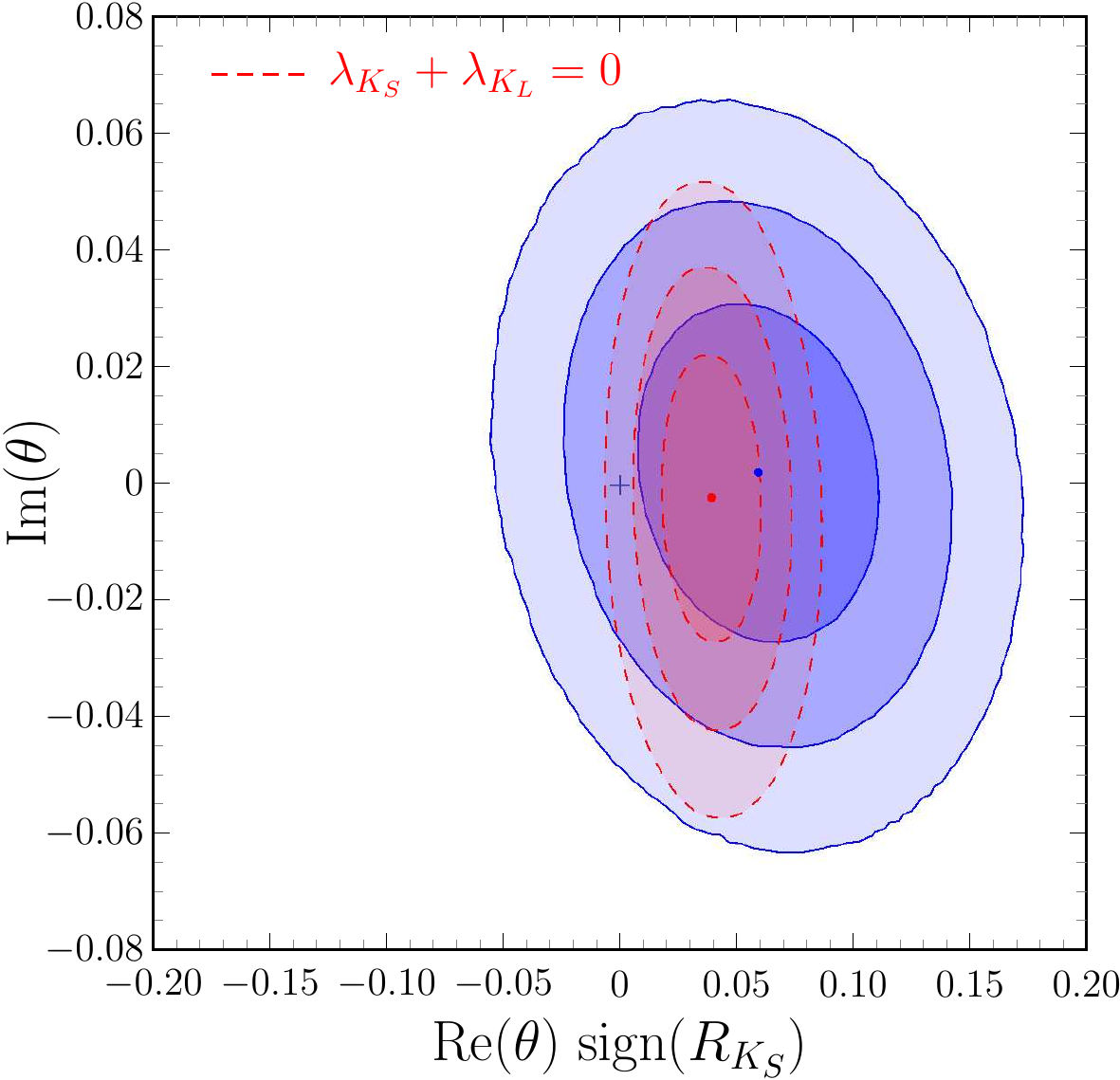}
\caption{$\im{\theta}$ vs. $\re{\theta}$sign$(\R{\KS})$ in the full fit (blue regions, solid contours), and in the fit with $\lambda_{\KS}+\lambda_{\KL}=0$ (red regions, dashed contours); dark to light regions correspond, respectively, to two-dimensional 68\%, 95\% and 99\% C.L. here and in all plots in the following.\label{FIG:ReImTheta}}
\end{center}
\end{figure}

With the information on the single $\FCOCc{\ell^\pm}{\KSL}$ and $\FCOSc{\ell^\pm}{\KSL}$ coefficients provided by the BaBar collaboration in \cite{Lees:2012uka}, including full covariance information and separate statistical and systematic uncertainties, supplemented with information on $\abs{q/p}$, for which we use \cite{PDG:2014} (obtained without assuming CPT invariance in the \BBdmix\ mixing)
\begin{equation}
\label{eq:qp:EXP}
\ABS{\frac{q}{p}}=1+\left(0.5\pm 1.1\right)\times 10^{-3}\,,
\end{equation}
we perform a fit in terms of the set of parameters $\{\re{\theta},\im{\theta},\delta,\rho,\beta,\epsilon_\rho,\epsilon_\beta\}$ (see \refEQ{eq:Lambdas:Param:01}). Furthermore, we can also address a more restricted situation where no wrong flavour decays (i.e. $\Delta F=\Delta Q$) are allowed in $\Bd,\Bdb\to J/\Psi \KSL$, that is imposing $\lambda_{\KS}+\lambda_{\KL}=0$: in terms of the previous set of parameters, that means setting $\rho=1$ and $\epsilon_\beta=0$. All the results shown in the following are obtained from a standard frequentist likelihood analysis. An additional bayesian analysis has also been performed with simple flat priors for the basic parameters, yielding almost identical results. Starting with the CPT violating $\theta$ parameter, the results that follow from these fits,
\begin{multline}
\label{eq:theta:fits:01}
\left\{
\begin{matrix}\re{\theta}=\pm(5.92\pm 3.03)\times 10^{-2}\\ \im{\theta}=(0.22\pm 1.90)\times 10^{-2}
\end{matrix}
\right\}
\\
\text{and}\
\left\{
\begin{matrix}
\re{\theta}=\pm(3.92\pm 1.43)\times 10^{-2}\\ \im{\theta}=(-0.22\pm 1.64)\times 10^{-2}
\end{matrix}
\right\}
\ \text{with}\ \lambda_{\KS}+\lambda_{\KL}=0,
\end{multline}
 improve significantly on the uncertainty of the real part quoted by the Particle Data Group (PDG) in \cite{PDG:2014}, based on BaBar \cite{Aubert:2004xga,Aubert:2006nf} and Belle \cite{Higuchi:2012kx} results (the PDG uses $z$ for our parameter $\theta$):
\begin{equation}
\label{eq:PDG:theta}
\re{\theta}_{\rm PDF}=\pm(1.9\pm 3.7\pm 3.3)\times 10^{-2}\,,\quad \im{\theta}_{\rm PDF}=(-0.8\pm 0.4)\times 10^{-2}\,.
\end{equation}
The sign ambiguity for $\re{\theta}$ in \refEQS{eq:theta:fits:01} and \eqref{eq:PDG:theta} is associated to the sign of $\R{\KS}$ and $\R{\KL}$: in the different asymmetries in sections \ref{SEC:Babar} and \ref{SEC:Genuine}, expanded to linear order in $\theta$, $\re{\theta}$ and $\R{\KSL}$ only appear multiplied together. Equation \eqref{eq:theta:fits:01} indeed corresponds to
\begin{multline}
\re{\theta}\,\text{sign}(\R{\KS})=(5.92\pm 3.03)\times 10^{-2},\\ \text{and}\ \re{\theta}\,\text{sign}(\R{\KS})=(3.92\pm 1.43)\times 10^{-2}\ \text{with}\ \lambda_{\KS}+\lambda_{\KL}=0.
\end{multline}
Following \refEQ{eq:H:params:01} with $\Delta\Gamma=0$, equation \eqref{eq:theta:fits:01} yields
\begin{multline}
\label{eq:theta:fits:02}
\left\{
\begin{matrix}\mathbf{M}_{22}-\mathbf{M}_{11}=\pm (2.0\pm 1.0)\\ \mathbf{\Gamma}_{22}-\mathbf{\Gamma}_{11}=-0.1\pm 1.3
\end{matrix}
\right\}10^{-5}\text{eV}\\
\text{and}\
\left\{
\begin{matrix}
\mathbf{M}_{22}-\mathbf{M}_{11}=\pm (1.3\pm 0.5)\\ \mathbf{\Gamma}_{22}-\mathbf{\Gamma}_{11}=0.1\pm 1.1
\end{matrix}
\right\}10^{-5}\text{eV}
\ \text{with}\ \lambda_{\KS}+\lambda_{\KL}=0.
\end{multline}
Figure \ref{FIG:ReImTheta} shows the result of the fit for the imaginary vs. real part of $\theta$.

Table \ref{TAB:fit:FULL} collects the results of the global fit to the data \cite{Lees:2012uka}, while the results in table \ref{TAB:fit:NWS} correspond to the fit with the additional assumption $\lambda_{\KS}+\lambda_{\KL}=0$. For completeness, the $\C{\KSL}$, $\S{\KSL}$ and $\R{\KSL}$ coefficients (see \refEQ{eq:MixingDecay:01}) are also displayed. Besides the basic parameters, BaBar asymmetries and genuine asymmetry coefficients are also shown, including separate values of the genuine T-reverse and fake contributions.

It has to be stressed that we do not observe any significant deviation of \refEQ{eq:Condition:MR:T:00}. This result confirms the goodness in the selection of the channels in order to constrain the T and CPT asymmetries. We also observe compatibility with the assumption of no wrong flavour decays in the CP final decay channel (results in Table \ref{TAB:fit:NWS}).

\begin{table}[h]
\begin{center}
\begin{tabular}{|l|c||l|c|}
\cline{1-4}
\multicolumn{4}{|c|}{WWA Parameters}\\ \hline
$\re{\theta}$ & $\pm(5.92\pm 3.03)10^{-2}$ & $\im{\theta}$ & $(0.22\pm 1.90)10^{-2}$ \\ \hline
$\rho$ & $1.021\pm 0.032$ & $\beta$ & $0.380\pm 0.020$\\ \hline
$\epsilon_\rho$ & $-0.023\pm 0.013$ & $\epsilon_\beta$ & $0.013\pm 0.040$\\ \hline
$\S{\KS}$ & $-0.679\pm 0.022$ & $\R{\KS}$ & $\pm(0.734\pm 0.020) $\\ \hline
$\C{\KS}$ & $\left(9.4\pm 3.22\right) 10^{-4}$\\ \hline
$\S{\KS}+\S{\KL}$ & $(1.9\pm 4.5)10^{-2}$ & $\R{\KS}+\R{\KL}$ & $(-1.9\pm 3.9)10^{-2}$\\ \hline
$\C{\KS}-\C{\KL}$ & $(-4.3\pm 6.0)10^{-2}$\\ \cline{1-2}\cline{1-2}\hline
\multicolumn{4}{|c|}{BaBar Asymmetries}\\ \hline
$\DST$ & $-1.317\pm 0.050$ & $\DSCP$ & $-1.360\pm 0.038$\\ \hline
$\DSCPT$ & $(7.6\pm 4.8)10^{-2}$\\ \hline
$\DCT$ & $(4.7\pm 3.7)10^{-2}$ & $\DCCP$ & $(8.9\pm 3.2)10^{-2}$\\ \hline
$\DCCPT$ & $(4.4\pm 3.6)10^{-2}$\\ \hline
\multicolumn{2}{|c||}{Genuine T-reverse} & \multicolumn{2}{|c|}{Fake} \\ \hline
$\DST$ g. & $-1.318\pm 0.050$ & $\DST$ f. & $(0.9 \pm 2.0)10^{-3}$\\ \hline
$\DSCPT$ g. & $(5.6\pm 4.3)10^{-2}$ & $\DSCPT$ f. & $(1.9\pm 4.7)10^{-2}$\\ \hline
$\DCT$ g. & $(0.2\pm 2.5)10^{-2}$ & $\DCT$ f. & $(4.5\pm 2.6)10^{-2}$\\ \hline
$\DCCPT$ g. & $(8.9\pm 5.2)10^{-2}$ & $\DCCPT$ f. & $(-4.5\pm 6.2)10^{-2}$\\ \hline\hline
\multicolumn{4}{|c|}{Genuine Asymmetry Parameters}\\ \hline
$\DScT$ & $-0.687\pm 0.020$ & $\DScCP$ & $-0.680\pm 0.021$\\ \hline
$\DScCPT$ & $(0.7\pm 2.0)10^{-2}$\\ \hline
$\DCcT$ & $(2.4\pm 2.0)10^{-2}$ & $\DCcCP$ & $(4.4\pm 1.6)10^{-2}$\\ \hline
$\DCcCPT$ & $(2.3\pm 1.8)10^{-2}$\\ \hline
$\DChT$ & $(-0.2\pm 1.4)10^{-2}$ & $\DChCP$ & $(-4.3\pm 2.5)10^{-2}$\\ \hline
$\DChCPT$ & $(-4.4\pm 2.6)10^{-2}$\\ \hline
\multicolumn{2}{|c||}{Genuine T-reverse} & \multicolumn{2}{|c|}{Fake} \\ \hline
$\DScT$ g.& $-0.687\pm 0.020$ & $\DScT$ f. & $(0.4\pm 1.2)10^{-3}$\\ \hline
$\DScCPT$ g.& $(-0.2\pm 1.9)10^{-2}$ & $\DScCPT$ f. & $(1.0\pm 2.4)10^{-2}$\\ \hline
$\DCcT$ g.& $(0.1\pm 1.4)10^{-2}$ & $\DCcT$ f. & $(2.3\pm 1.3)10^{-2}$\\ \hline
$\DCcCPT$ g.& $(2.4\pm 2.6)10^{-2}$ & $\DCcCPT$ f. & $(-2.1\pm 3.2)10^{-2}$\\ \hline
$\DChT$ g.& $(-0.1\pm 1.4)10^{-2}$ & $\DChT$ f. & $(-0.5\pm 1.6)10^{-3}$\\ \hline
$\DChCPT$ g.& $(-4.4\pm 2.7)10^{-2}$ & $\DChCPT$ f. & $(0.6\pm 5.0)10^{-4}$\\ \hline
\end{tabular}
\caption{Global fit, summary of results.\label{TAB:fit:FULL}}
\end{center}
\end{table}

\begin{table}[h]
\begin{center}
\begin{tabular}{|l|c||l|c|}
\cline{1-4}
\multicolumn{4}{|c|}{WWA Parameters}\\ \hline
$\re{\theta}$ & $\pm (3.92\pm 1.43)10^{-2}$ & $\im{\theta}$ & $(-0.22\pm 1.64)10^{-2}$ \\ \hline
$\epsilon_\rho$ & $-0.021\pm 0.013$ & $\beta$ & $0.375\pm 0.016$\\ \hline
$\S{\KS}$ & $-0.682\pm 0.017$ & $\R{\KS}$ & $\pm(0.731\pm 0.016) $\\ \hline
$\C{\KS}$ & $(2.10\pm 1.31) 10^{-2}$\\ \cline{1-2}
\hline
\multicolumn{4}{|c|}{BaBar Asymmetries}\\ \hline
$\DST$ & $-1.326\pm 0.033$ & $\DSCP$ & $-1.362\pm 0.0358$\\ \hline
$\DSCPT$ & $(4.1\pm 2.3)10^{-2}$\\ \hline
$\DCT$ & $(3.8\pm 3.4)10^{-2}$ & $\DCCP$ & $0.100\pm 0.029$\\ \hline
$\DCCPT$ & $(5.3\pm 2.9)10^{-2}$\\ \hline
\multicolumn{2}{|c||}{Genuine T-reverse} & \multicolumn{2}{|c|}{Fake} \\ \hline
$\DST$ g. & $-1.326\pm 0.033$ & $\DST$ f. & $(1.9\pm \begin{smallmatrix}+ 10.0\\ -7.5\end{smallmatrix})10^{-4}$\\ \hline
$\DSCPT$ g. & $(4.1\pm 2.3)10^{-2}$ & $\DSCPT$ f. & $(-1.1\pm 8.0)10^{-4}$\\ \hline
$\DCT$ g. & $(0.4\pm 2.2)10^{-2}$ & $\DCT$ f. & $(4.2\pm 2.6)10^{-2}$\\ \hline
$\DCCPT$ g. & $(5.4\pm 2.9)10^{-2}$ & $\DCCPT$ f. & $(-1.2\pm 1.0)10^{-3}$\\ \hline\hline
\multicolumn{4}{|c|}{Genuine Asymmetry Parameters}\\ \hline
$\DScT$ & $-0.682\pm 0.017$ & $\DScCP$ & $-0.680\pm 0.022$\\ \hline
$\DScCPT$ & $(0.2\pm 1.6)10^{-2}$\\ \hline
$\DCcT$ & $(2.0\pm 1.8)10^{-2}$ & $\DCcCP$ & $(5.0\pm 1.5)10^{-2}$\\ \hline
$\DCcCPT$ & $(2.7\pm 1.5)10^{-2}$\\ \hline
$\DChT$ & $(0.2\pm 1.2)10^{-2}$ & $\DChCP$ & $(-2.8\pm 1.0)10^{-2}$\\ \hline
$\DChCPT$ & $(-2.7\pm 1.5)10^{-2}$\\ \hline
\multicolumn{2}{|c||}{Genuine T-reverse} & \multicolumn{2}{|c|}{Fake} \\ \hline
$\DScT$ g.& $-0.682\pm 0.017$ & $\DScT$ f. & $(1.1\pm 5.1)10^{-4}$\\ \hline
$\DScCPT$ g.& $(0.2\pm 1.7)10^{-2}$ & $\DScCPT$ f. & $(-0.5\pm 4.4)10^{-4}$\\ \hline
$\DCcT$ g.& $(-0.2\pm 1.2)10^{-2}$ & $\DCcT$ f. & $(2.1\pm 1.3)10^{-2}$\\ \hline
$\DCcCPT$ g.& $(2.7\pm 1.5)10^{-2}$ & $\DCcCPT$ f. & $(0.6\pm 4.0)10^{-5}$\\ \hline
$\DChT$ g.& $(0.2\pm 1.2)10^{-2}$ & $\DChT$ f. & $(3.3\pm 4.0)10^{-5}$\\ \hline
$\DChCPT$ g.& $(-2.7\pm 1.5)10^{-2}$ & $\DChCPT$ f. & $(0.6\pm 2.0)10^{-5} $\\ \hline
\end{tabular}
\caption{Global fit with $\lambda_{\KS}+\lambda_{\KL}=0$, summary of results.\label{TAB:fit:NWS}}
\end{center}
\end{table}

\clearpage
\subsection{Selected results\label{sSEC:Fit:Various}}
\begin{figure}[h]
\begin{center}
\subfigure[$\DeltaS{}$.]{\includegraphics[width=0.3\textwidth]{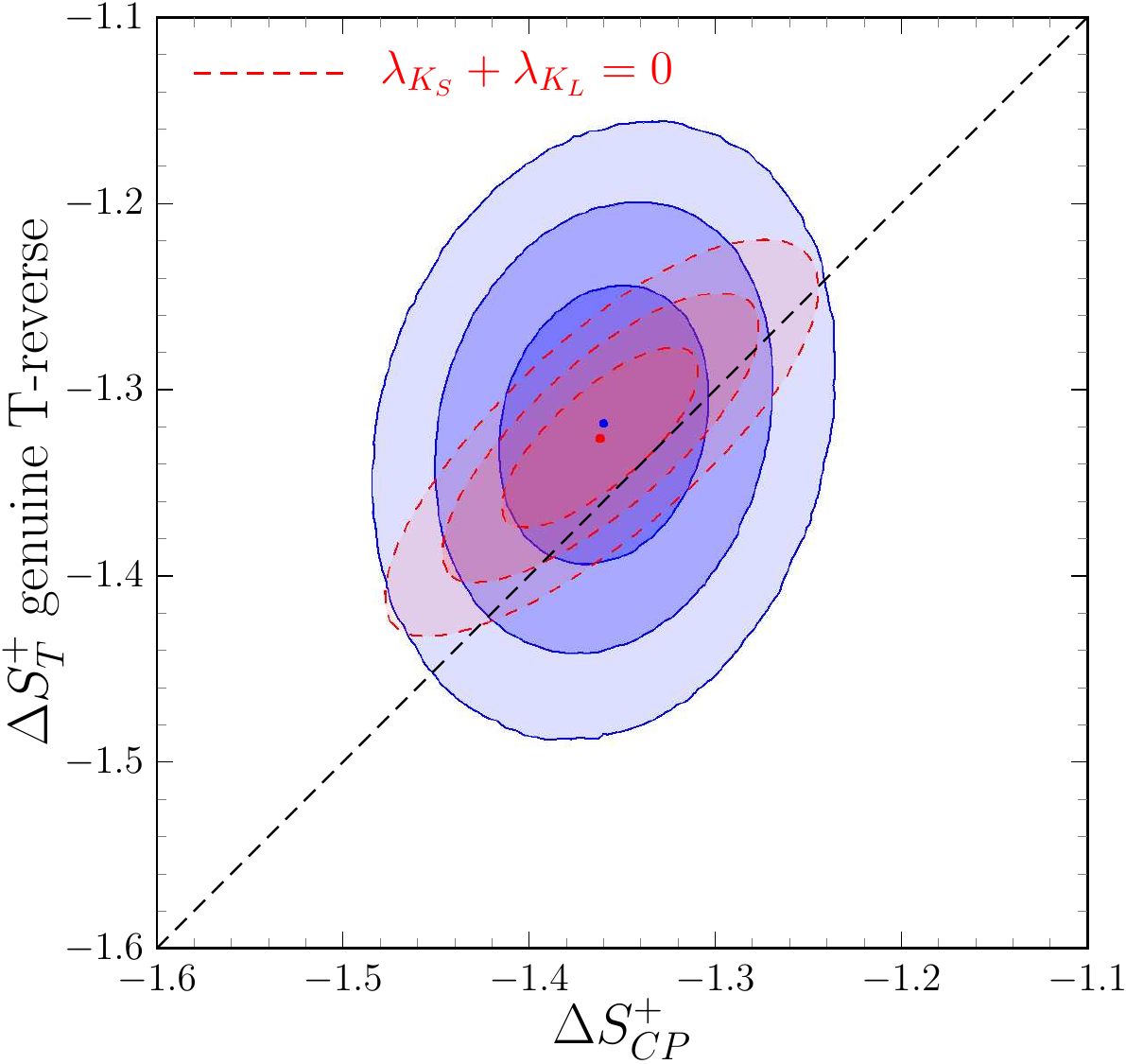}}\quad
\subfigure[$\DeltaSc{}$.]{\includegraphics[width=0.3\textwidth]{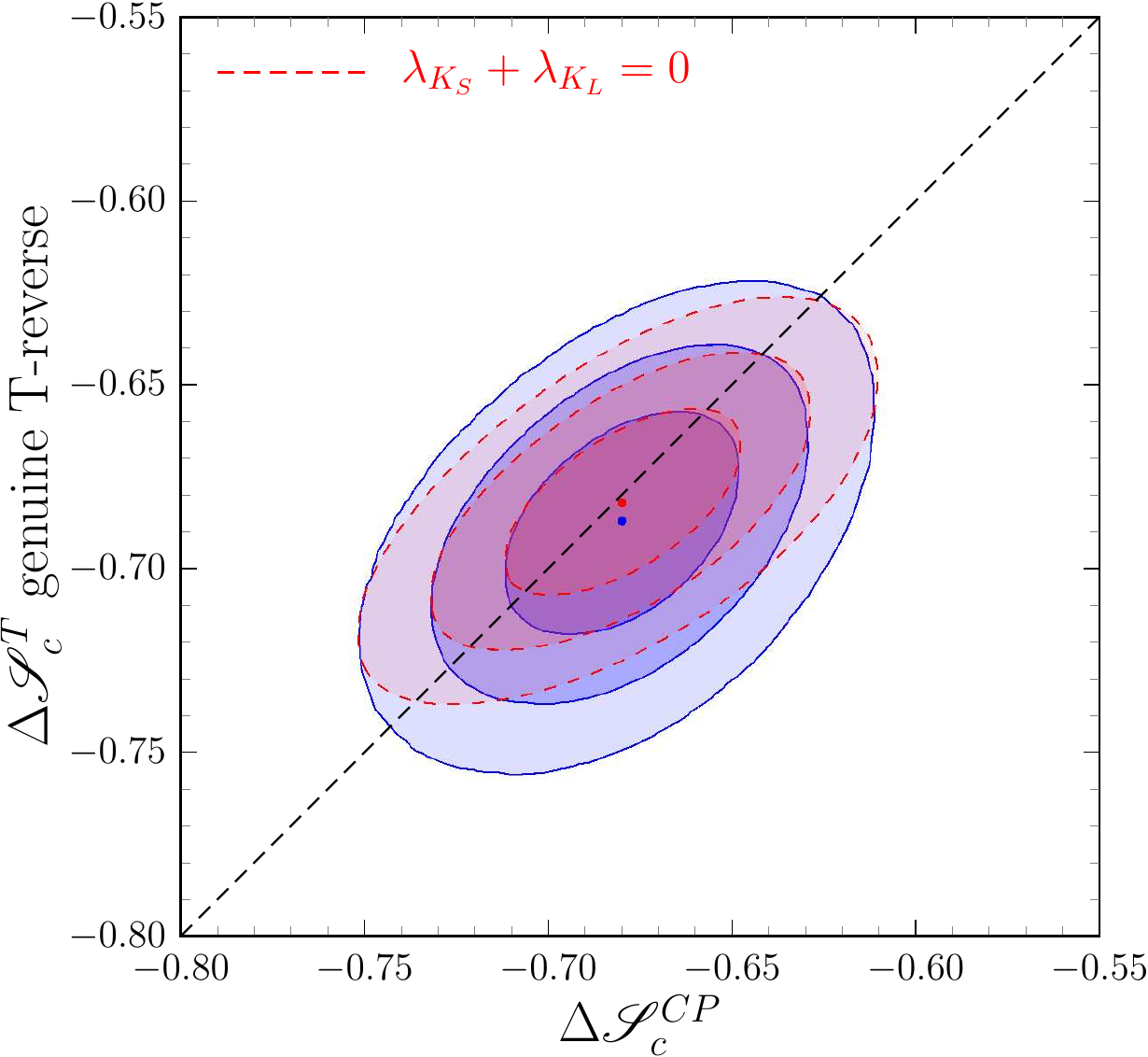}}\quad
\subfigure[$\DeltaCc{}$.\label{FIG:Babar:Ttrue-CP:Cc}]{\includegraphics[width=0.3\textwidth]{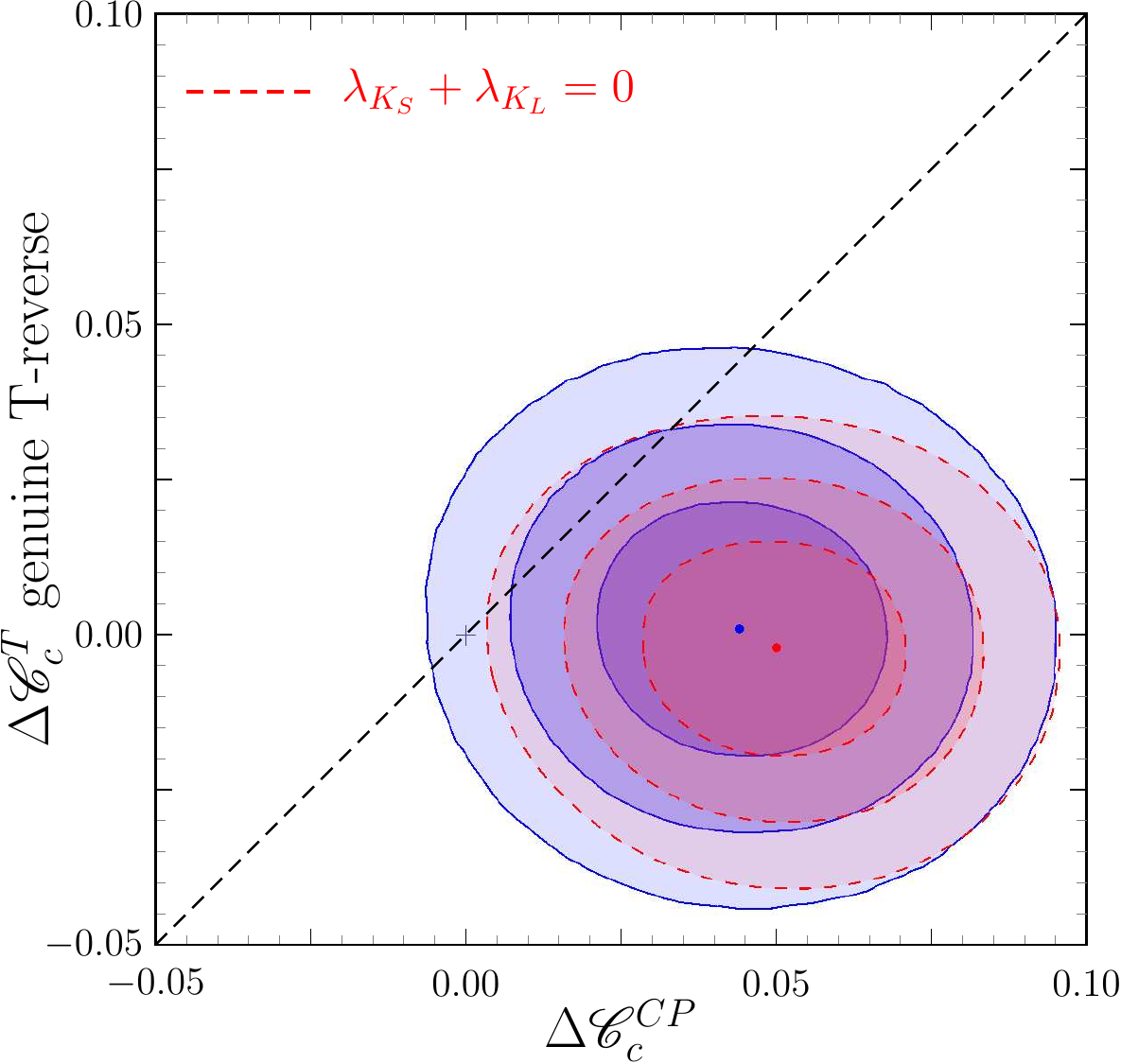}}
\caption{The difference between genuine T-reverse and CP asymmetries.}\label{FIG:Babar:Ttrue-CP}
\end{center}
\end{figure}

For the BaBar asymmetries we obtain, in the present analysis, $\DST=-1.317\pm 0.050$ and $\DST=-1.326\pm 0.033$ (asuming no wrong flavour decays). The remarkable improvement on the precision comes from imposing the WWA evolution, that includes symmetries like \refEQ{eq:IntensityCoefficients:04}. The CP counterpart is the asymmetry $\DSCP=-1.360\pm 0.038$. We now discuss the difference between the genuine T-reverse and CP asymmetry parameters.
 To illustrate this point, figure \ref{FIG:Babar:Ttrue-CP} shows true T-reverse asymmetries versus CP asymmetries for $\DeltaS{}$ and for the genuine asymmetry coefficients $\DeltaSc{}$ and $\DeltaCc{}$. The dashed diagonal line would correspond to strict equality among both observables.

\begin{figure}[h]
\begin{center}
\subfigure[$\DST$.\label{sFIG:TrueFake:DST}]{\includegraphics[width=0.3\textwidth]{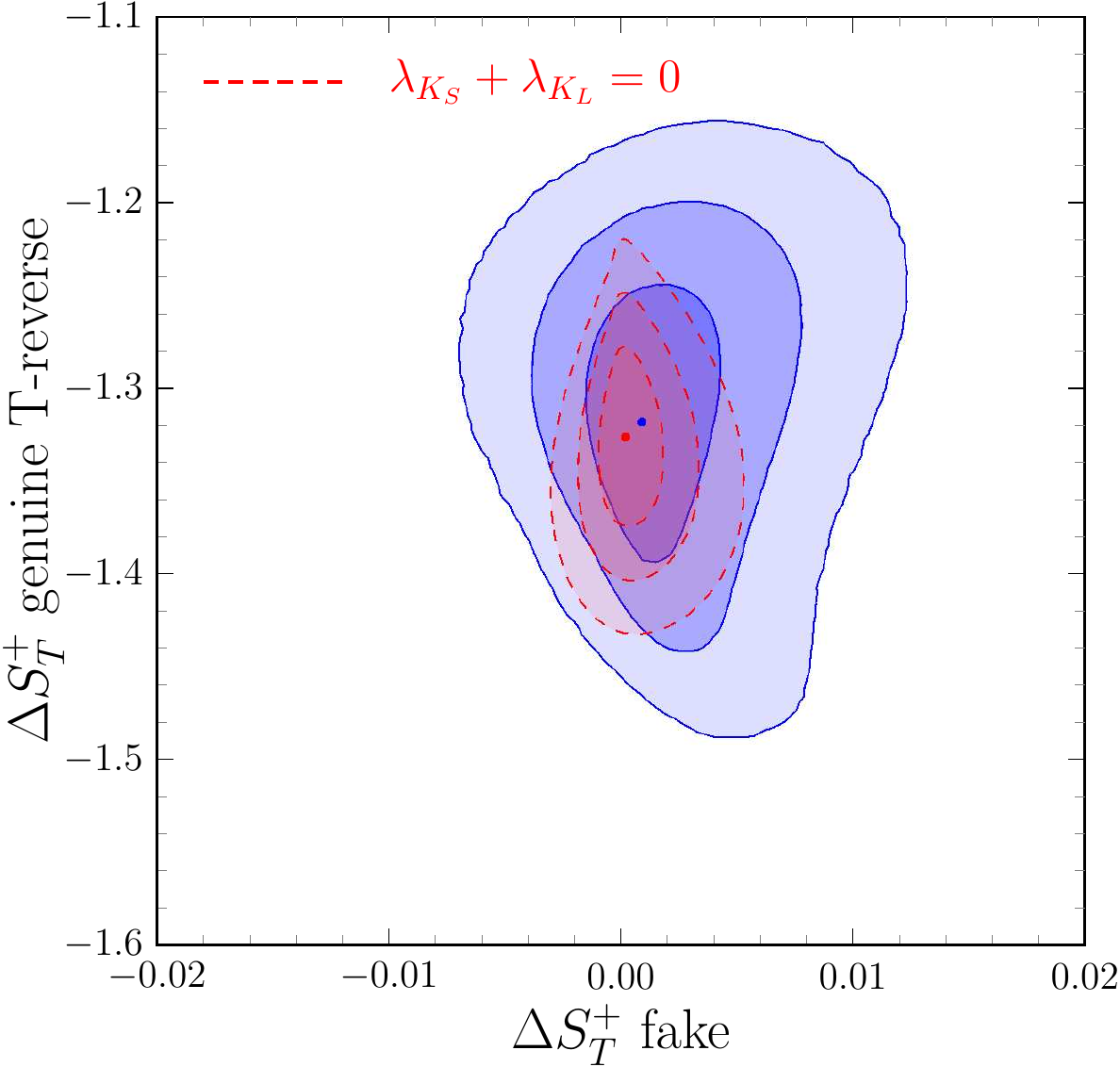}}\quad
\subfigure[$\DScT$.\label{sFIG:TrueFake:DScT}]{\includegraphics[width=0.3\textwidth]{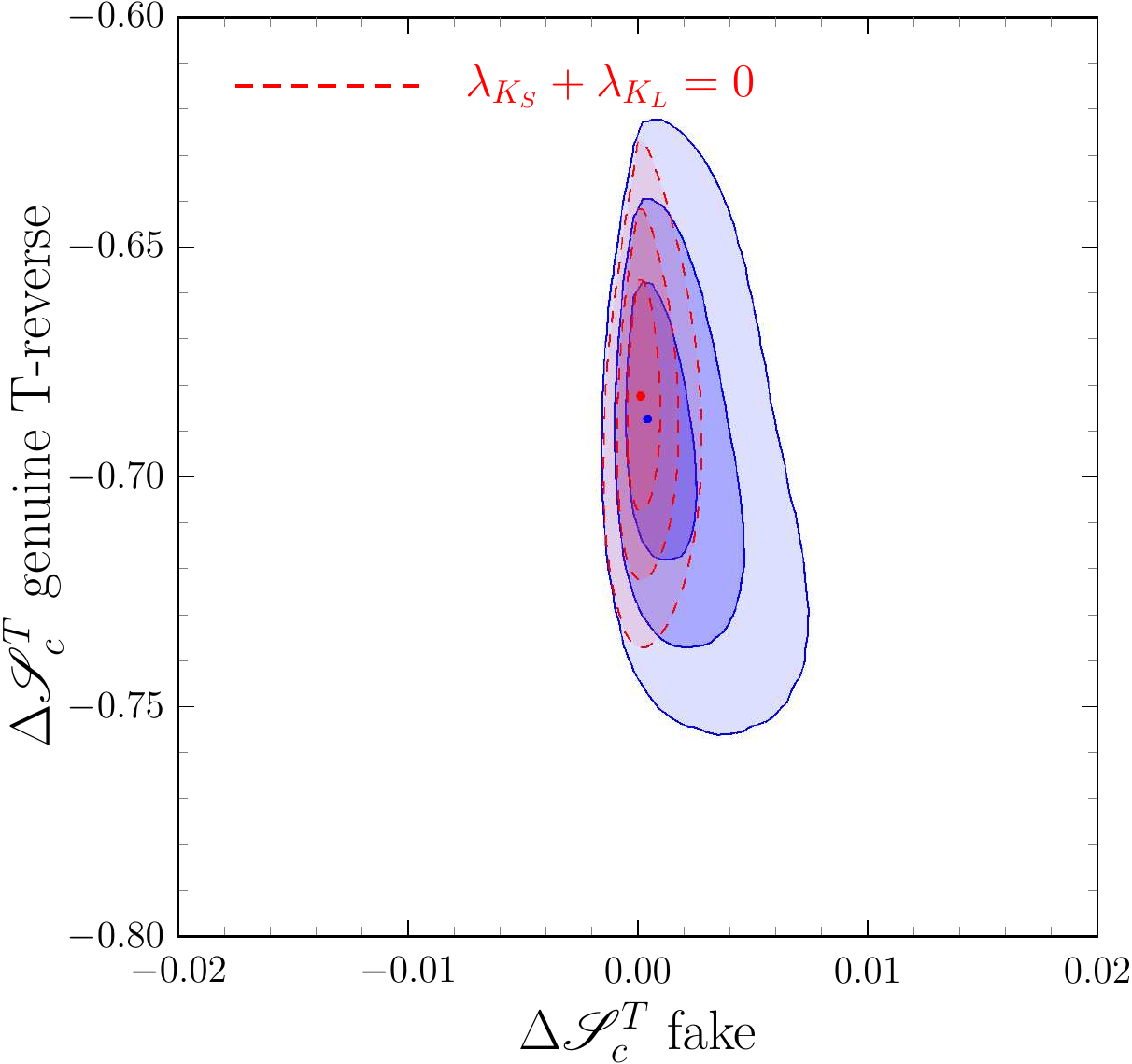}}\\
\subfigure[$\DChT$.\label{sFIG:TrueFake:DChT}]{\includegraphics[width=0.3\textwidth]{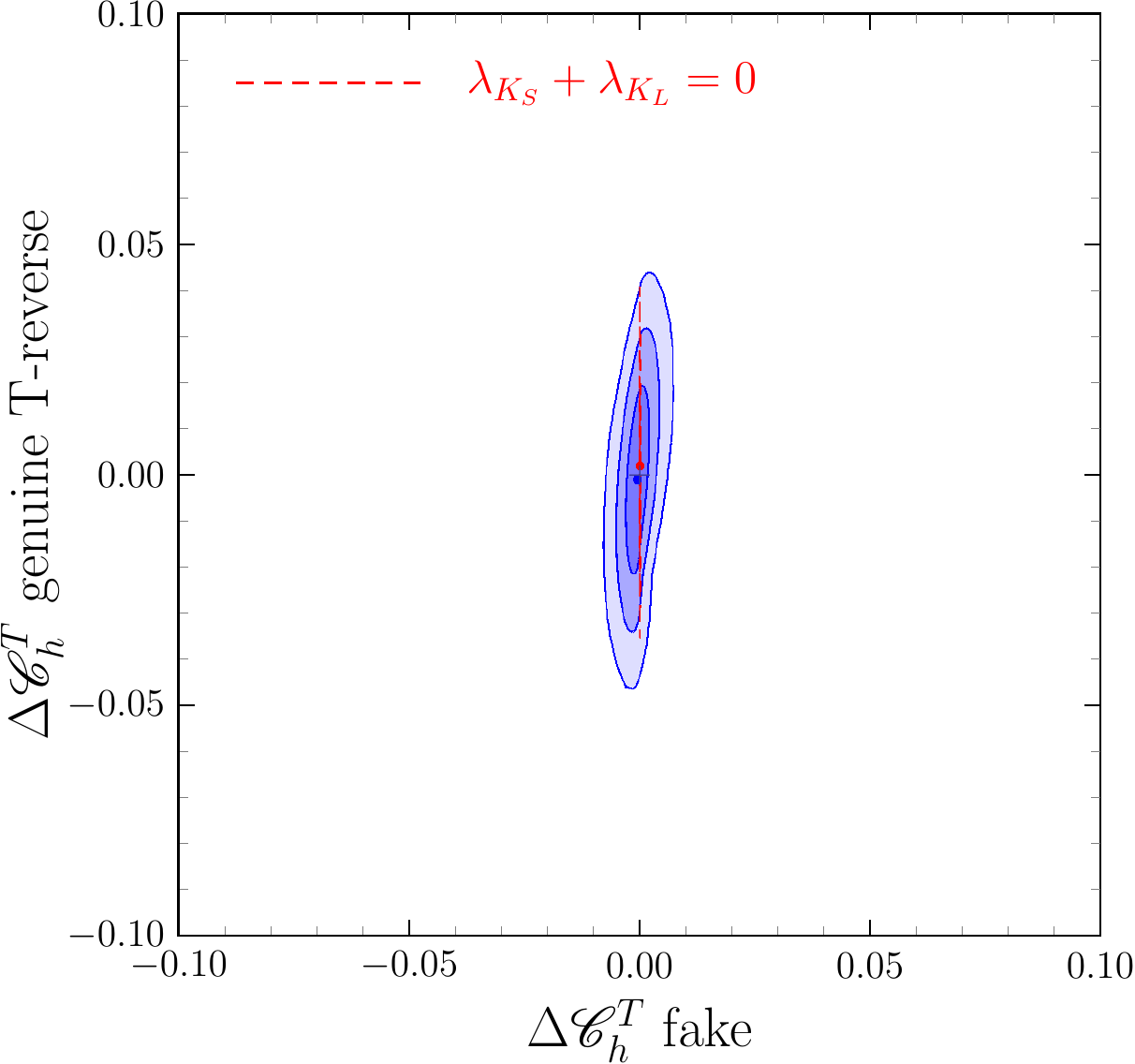}}\quad
\subfigure[$\DCcT$.\label{sFIG:TrueFake:DCcT}]{\includegraphics[width=0.3\textwidth]{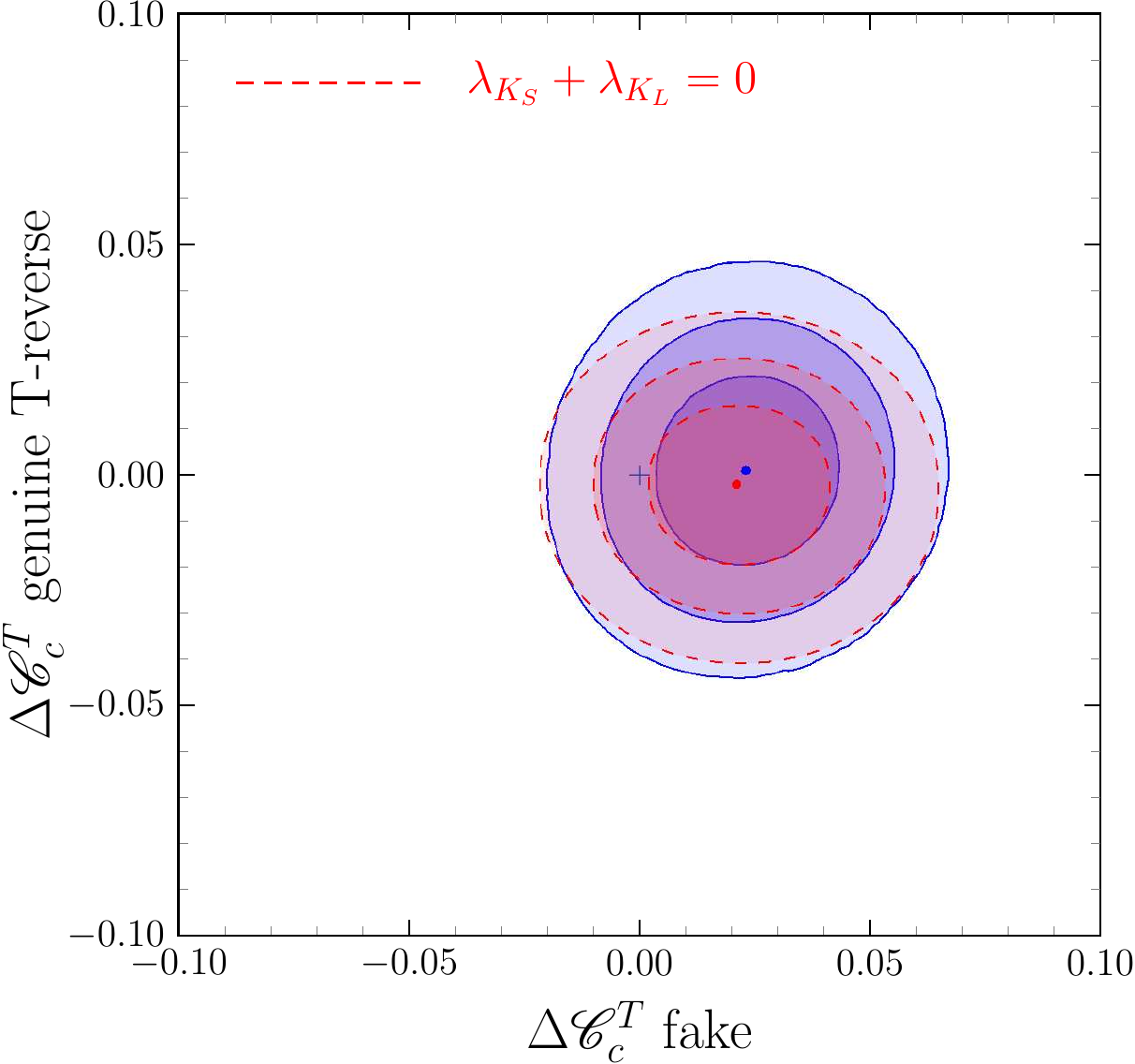}}
\caption{The separation of genuine T-reverse and fake contributions.}\label{FIG:TrueFake}
\end{center}
\end{figure}

In section \ref{SEC:TrueFake} we have shown that the genuine T-reverse and fake contributions to T and CPT asymmetries could be separated quantitatively. This is particularly relevant for the $\DST$ asymmetry since sizable fake contributions could have weakened the evidence for the time reversal violation observation independent of CP. In figure \ref{FIG:TrueFake} we show genuine T-reverse vs. fake contributions for $\DST$ and for the genuine asymmetry parameters $\DScT$, $\DCcT$ and $\DChT$: from figures \ref{sFIG:TrueFake:DST} and \ref{sFIG:TrueFake:DScT}, it is clear that the T-fake contributions to $\DST$ and $\DScT$ are below the percent level in accordance with expectations from the fit to $\rho$, $\epsilon_\rho$ and $\epsilon_\beta$. For $\DChT$ in figure \ref{sFIG:TrueFake:DChT}, although the fake contribution is below the $10^{-2}$ level while the genuine T-reverse one can reach the few percent level, there is no evidence of time reversal violation. For $\DCcT$ in figure \ref{sFIG:TrueFake:DCcT}, fake contributions might be as large as the genuine T-reverse ones and the same conclusion holds. It is to be noticed that, in all cases, there is no significant correlation among genuine T-reverse and fake contributions.

%

\begin{figure}[h]
\begin{center}
\subfigure[$\DChCP$.]{\includegraphics[width=0.3\textwidth]{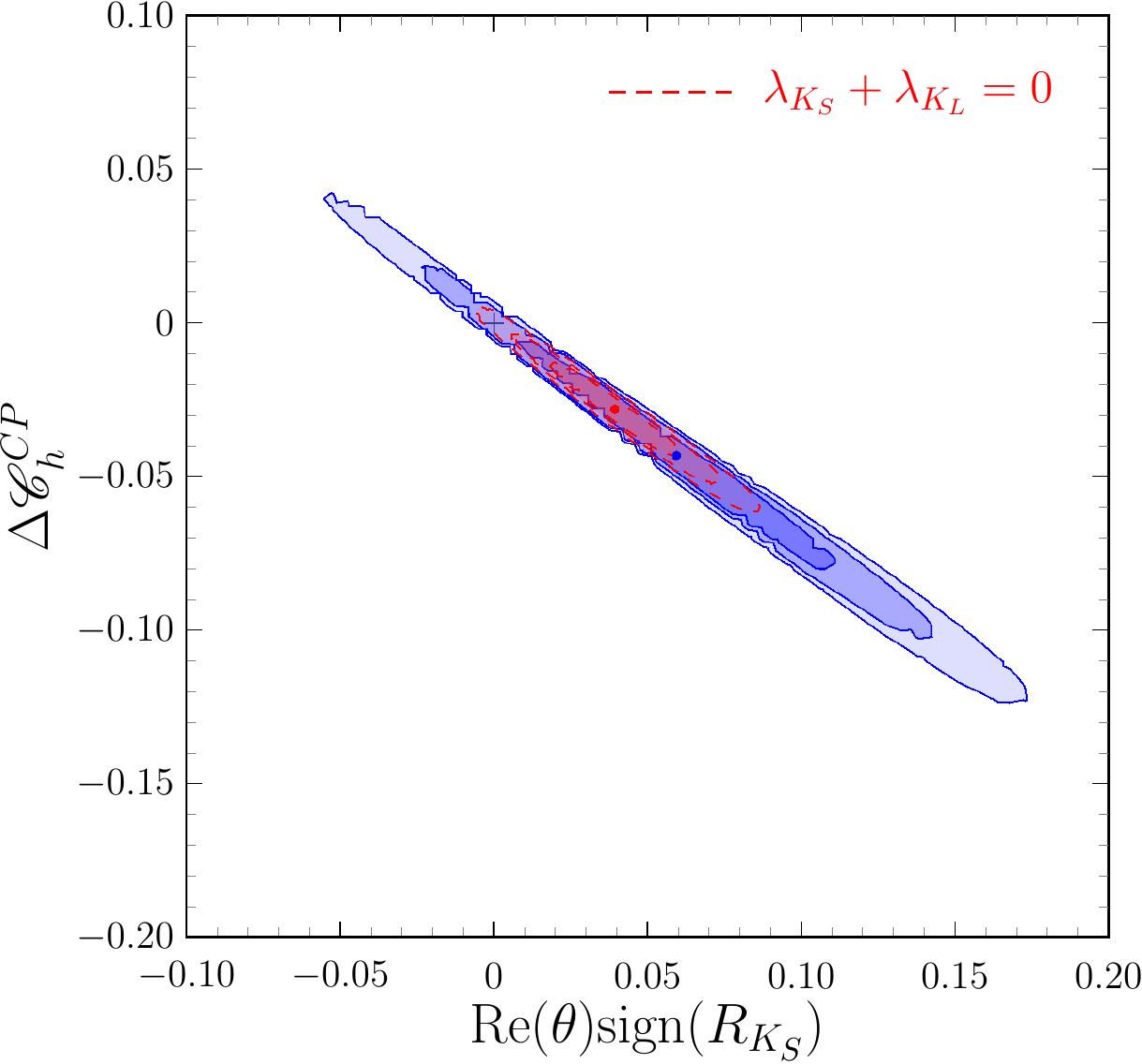}}\quad
\subfigure[$\DChCPT$.]{\includegraphics[width=0.3\textwidth]{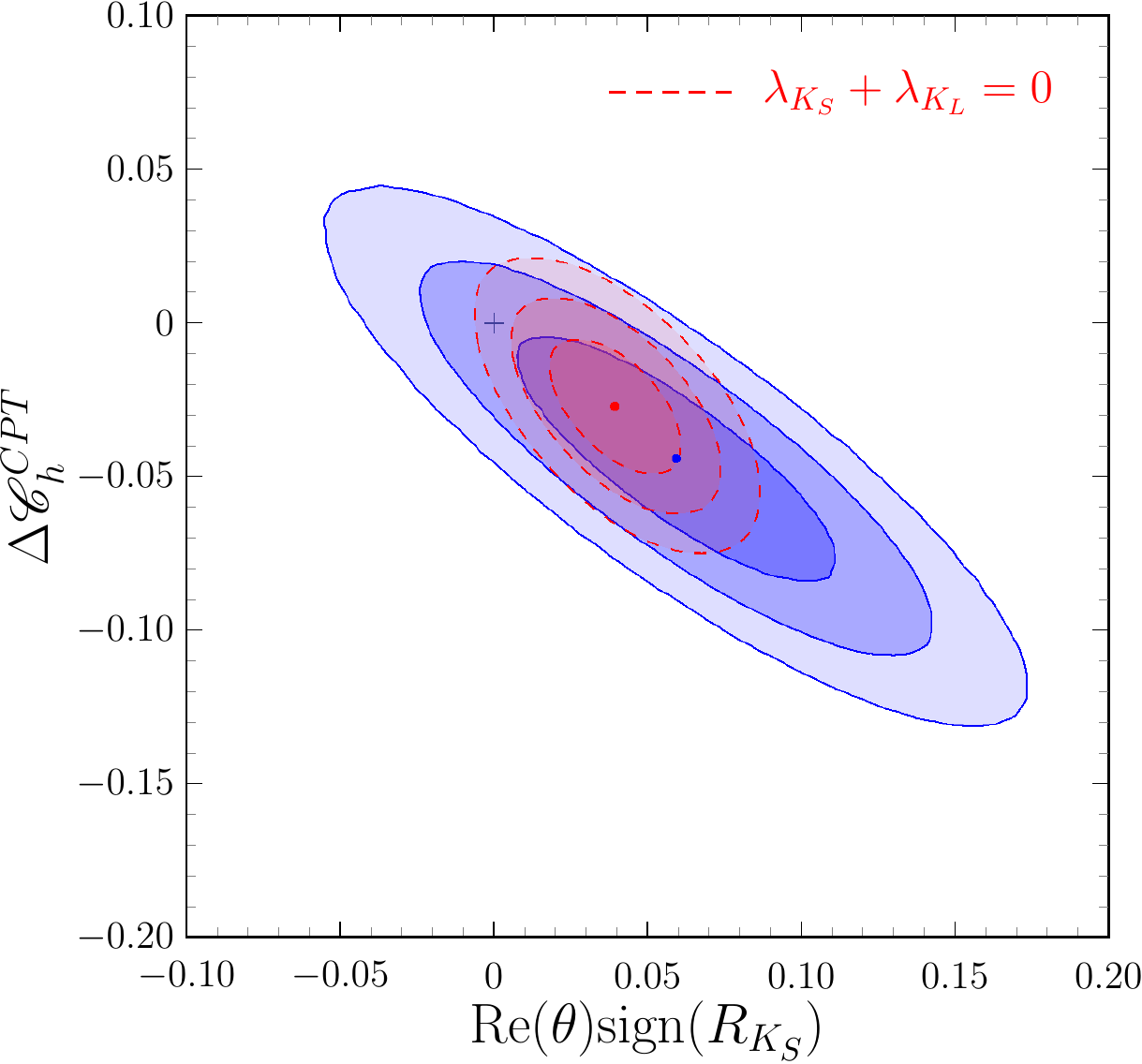}}\quad
\subfigure[Genuine T-reverse $\DChCPT$.]{\includegraphics[width=0.3\textwidth]{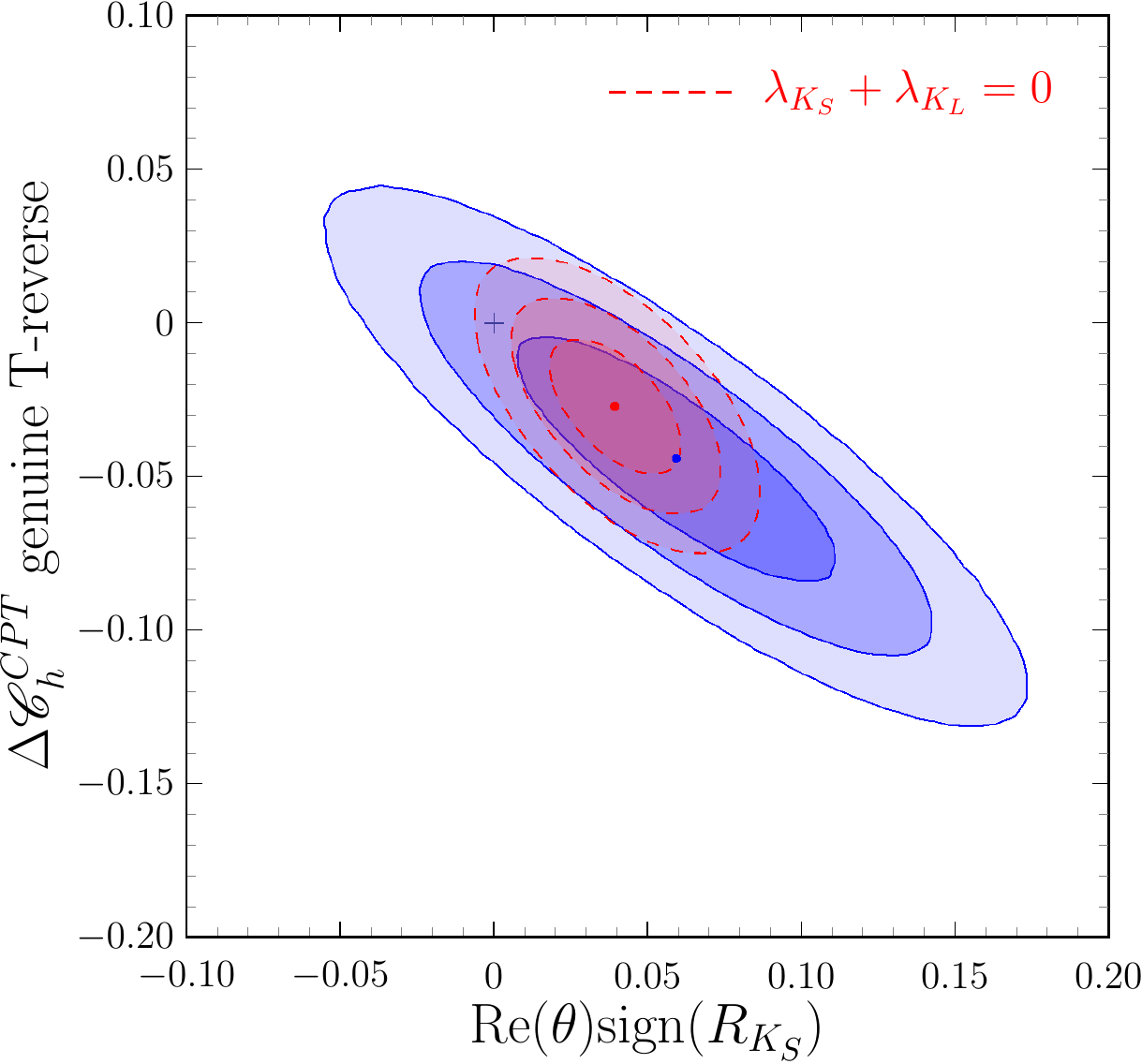}}\\
\subfigure[$\DCcCP$.]{\includegraphics[width=0.3\textwidth]{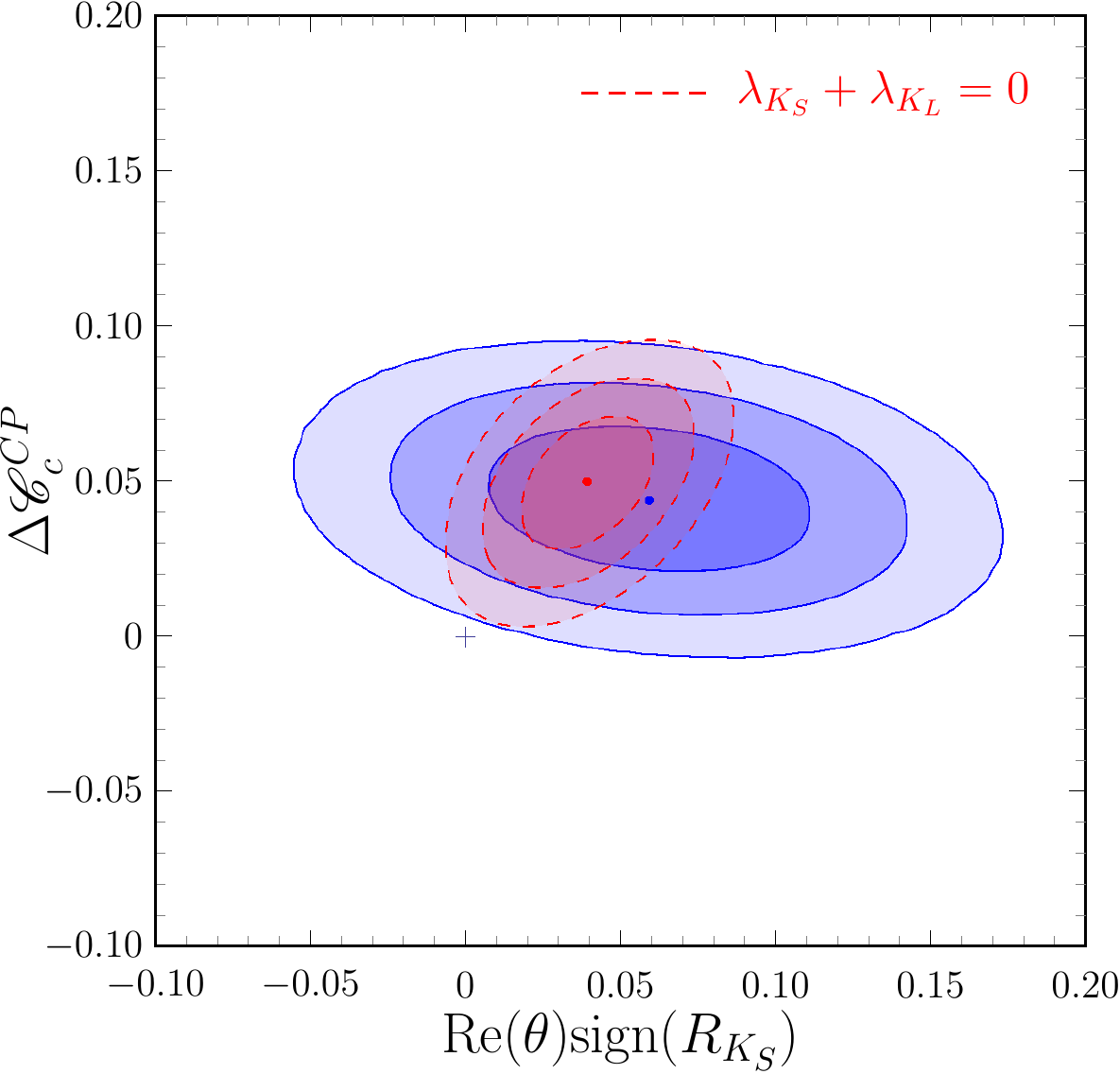}}\quad
\subfigure[$\DCcCPT$.]{\includegraphics[width=0.3\textwidth]{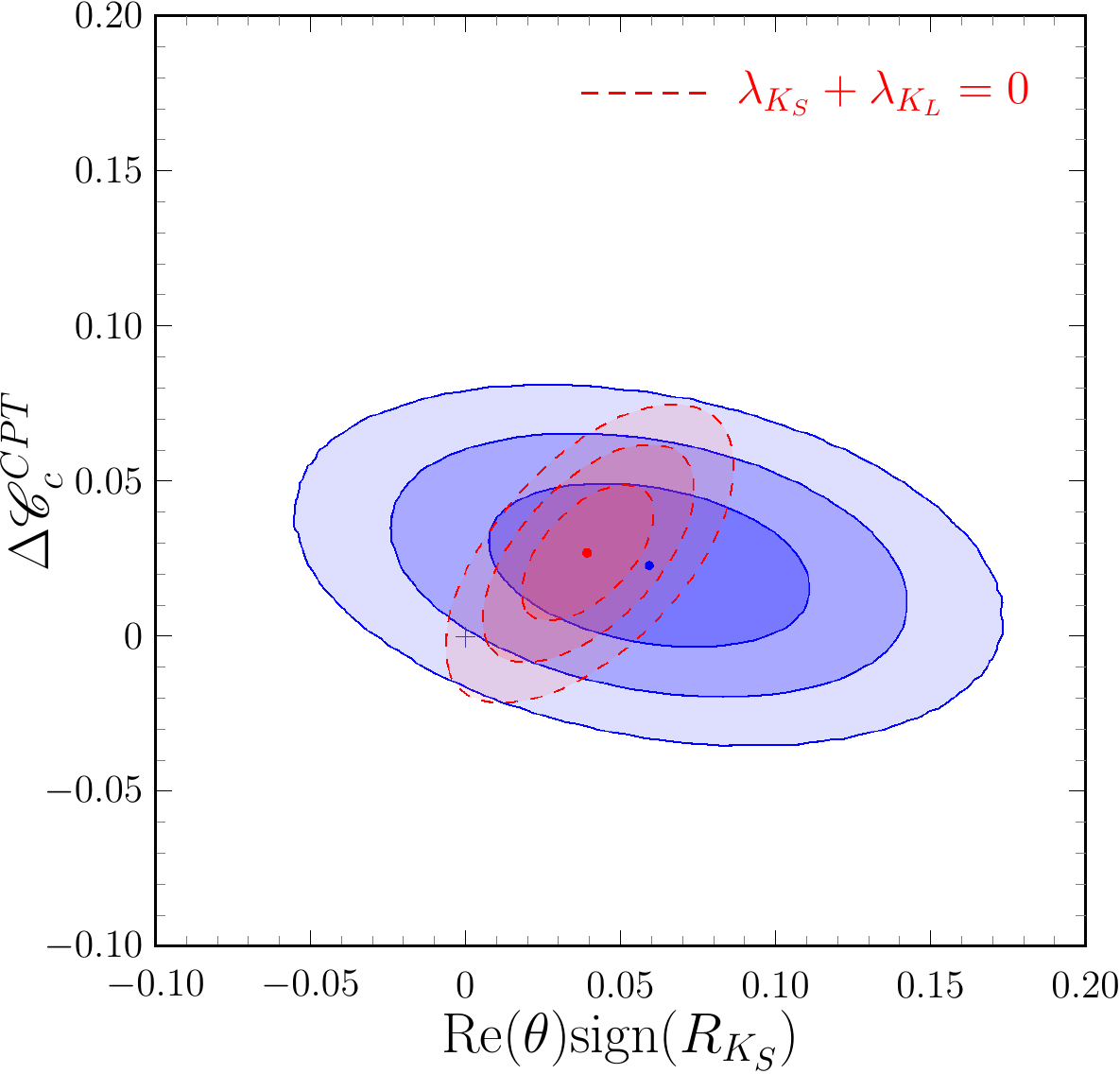}}\quad
\subfigure[Genuine T-reverse $\DCcCPT$.]{\includegraphics[width=0.3\textwidth]{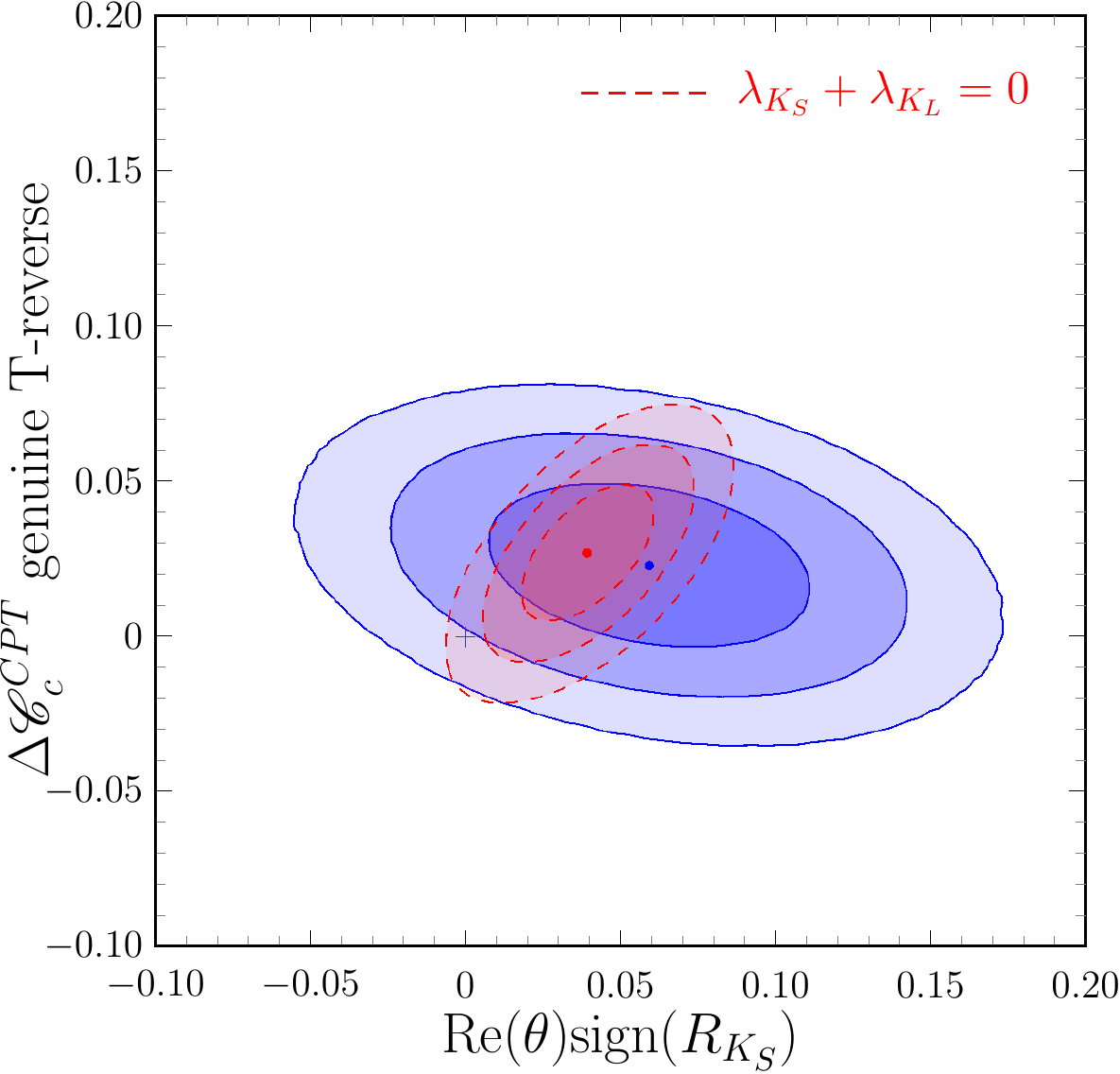}}
\caption{Correlations with $\re{\theta}$sign$(\R{\KS})$.}\label{FIG:ReThetaCorr}
\end{center}
\end{figure}

\begin{figure}[h]
\begin{center}
\subfigure[$\DChT$.]{\includegraphics[width=0.23\textwidth]{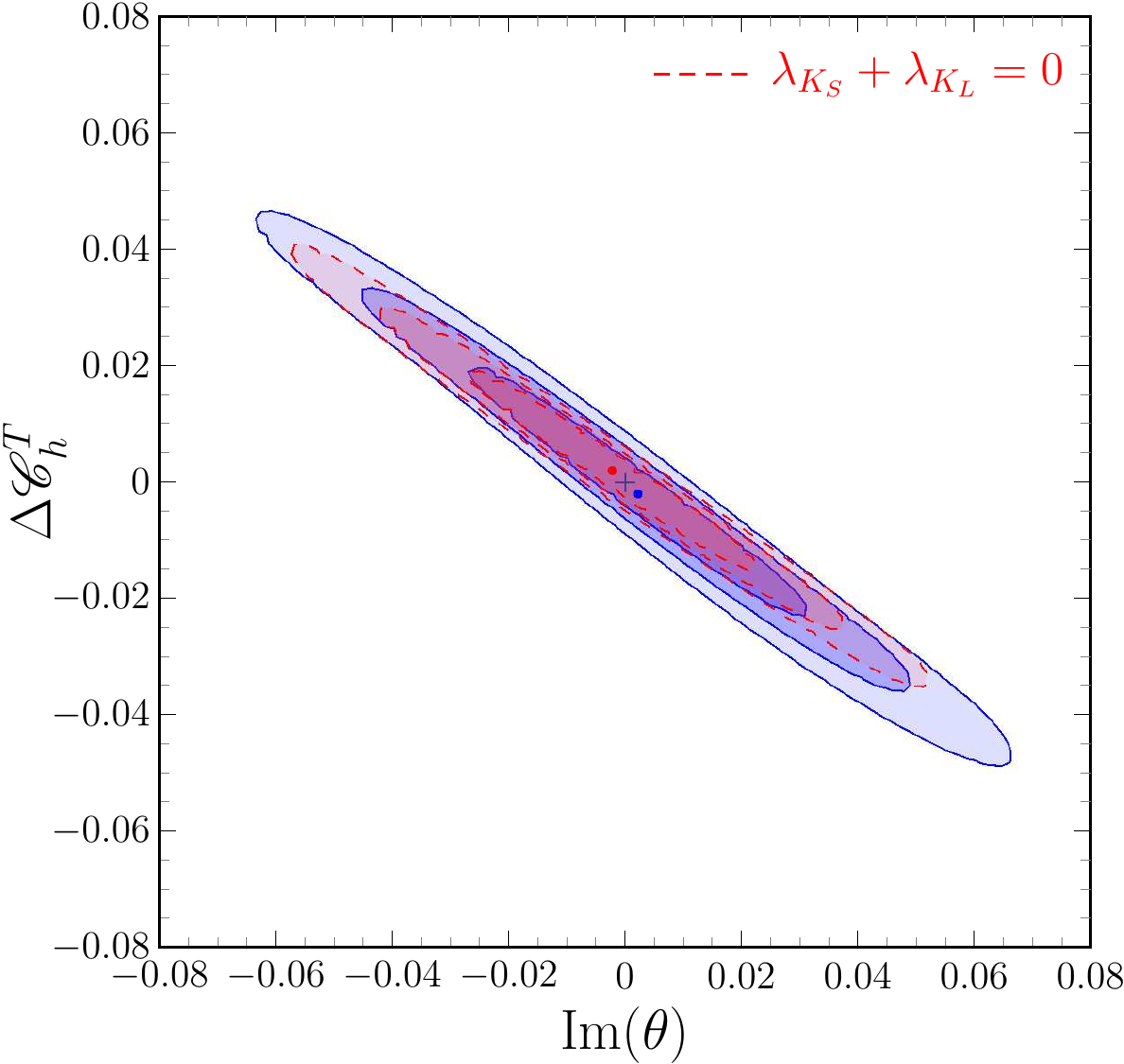}}\ 
\subfigure[Gen. T-rev. $\DChT$.]{\includegraphics[width=0.23\textwidth]{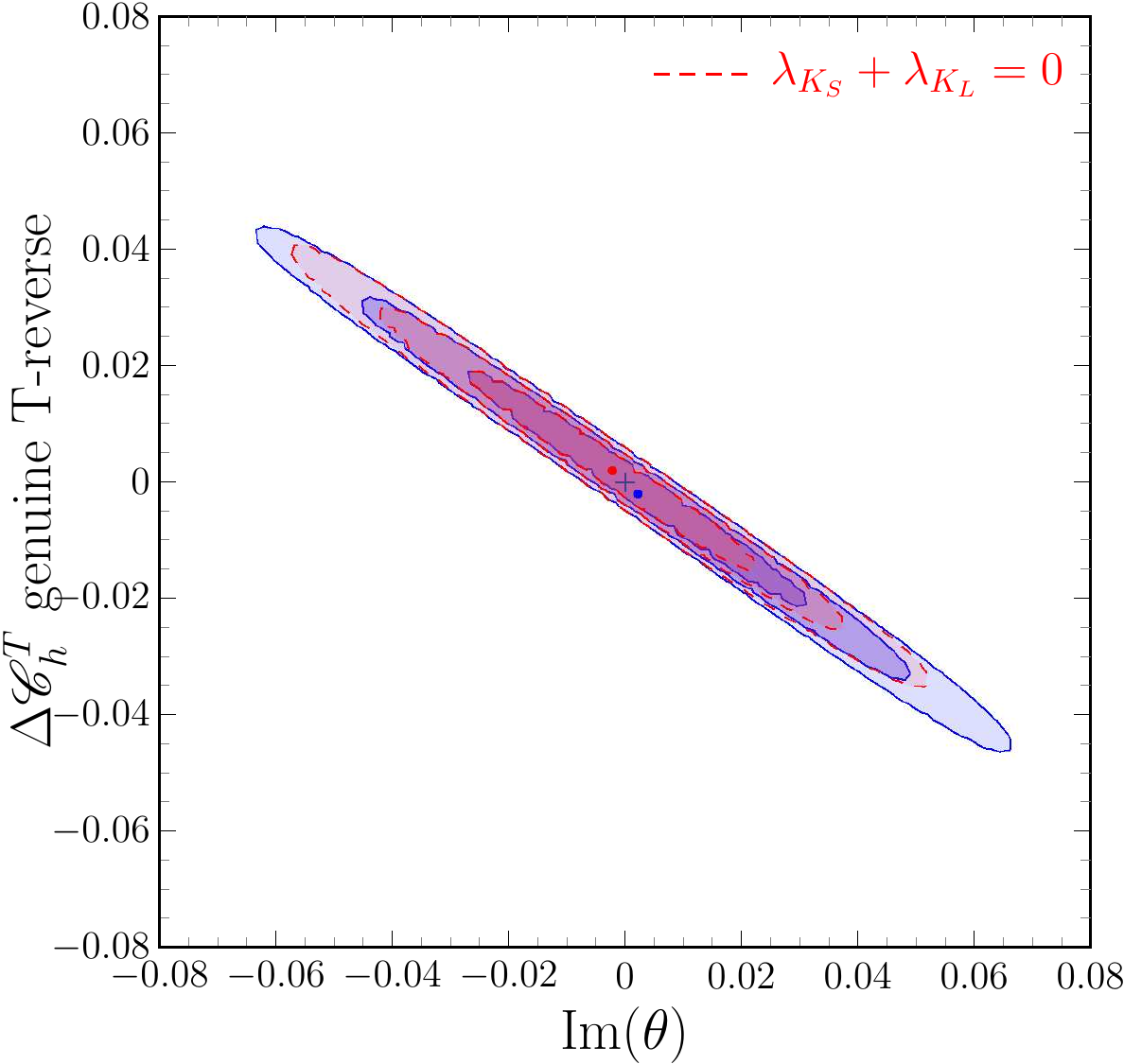}}\ 
\subfigure[$\DCcT$.]{\includegraphics[width=0.23\textwidth]{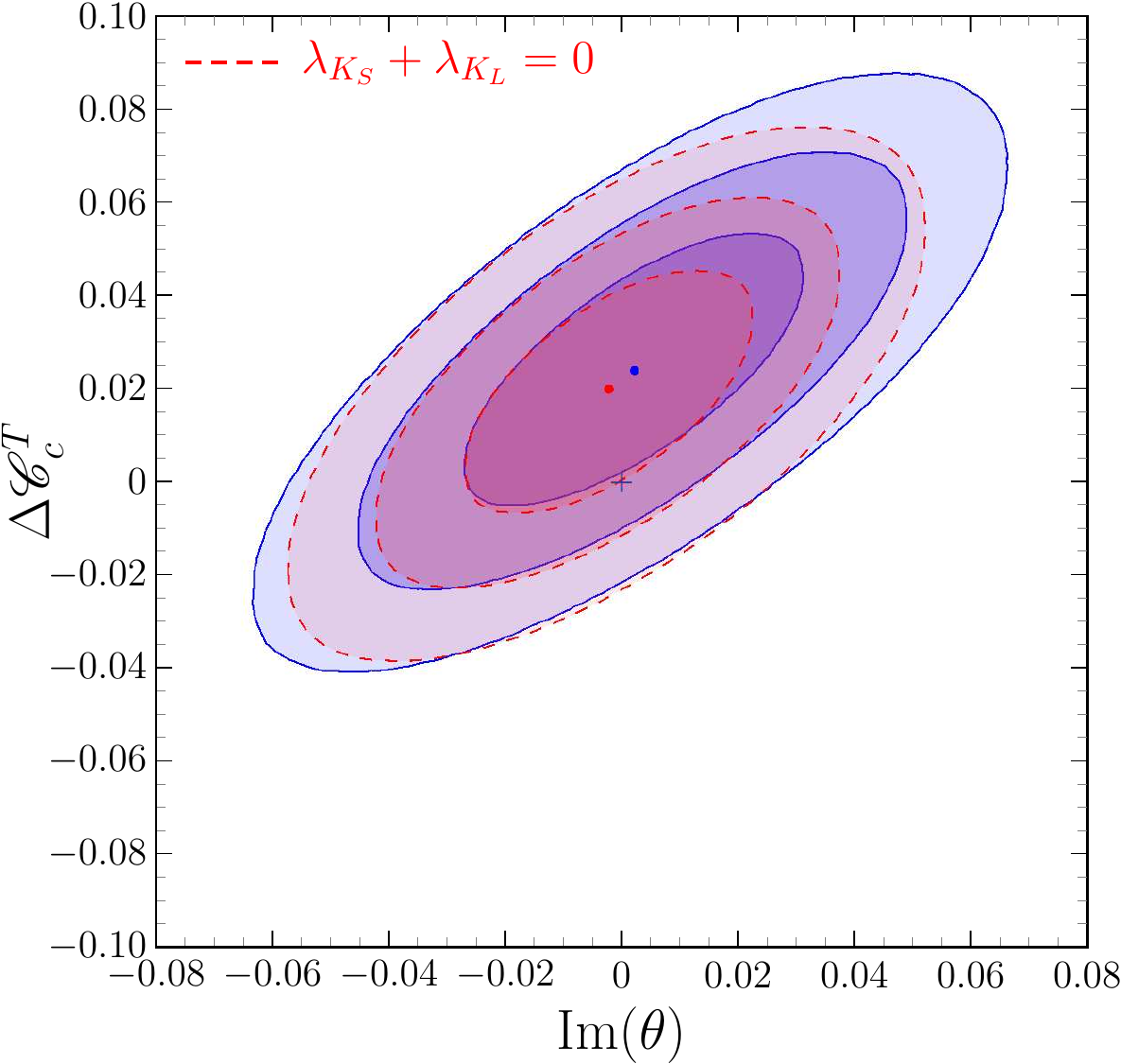}}\ 
\subfigure[Gen. T-rev. $\DCcT$.]{\includegraphics[width=0.23\textwidth]{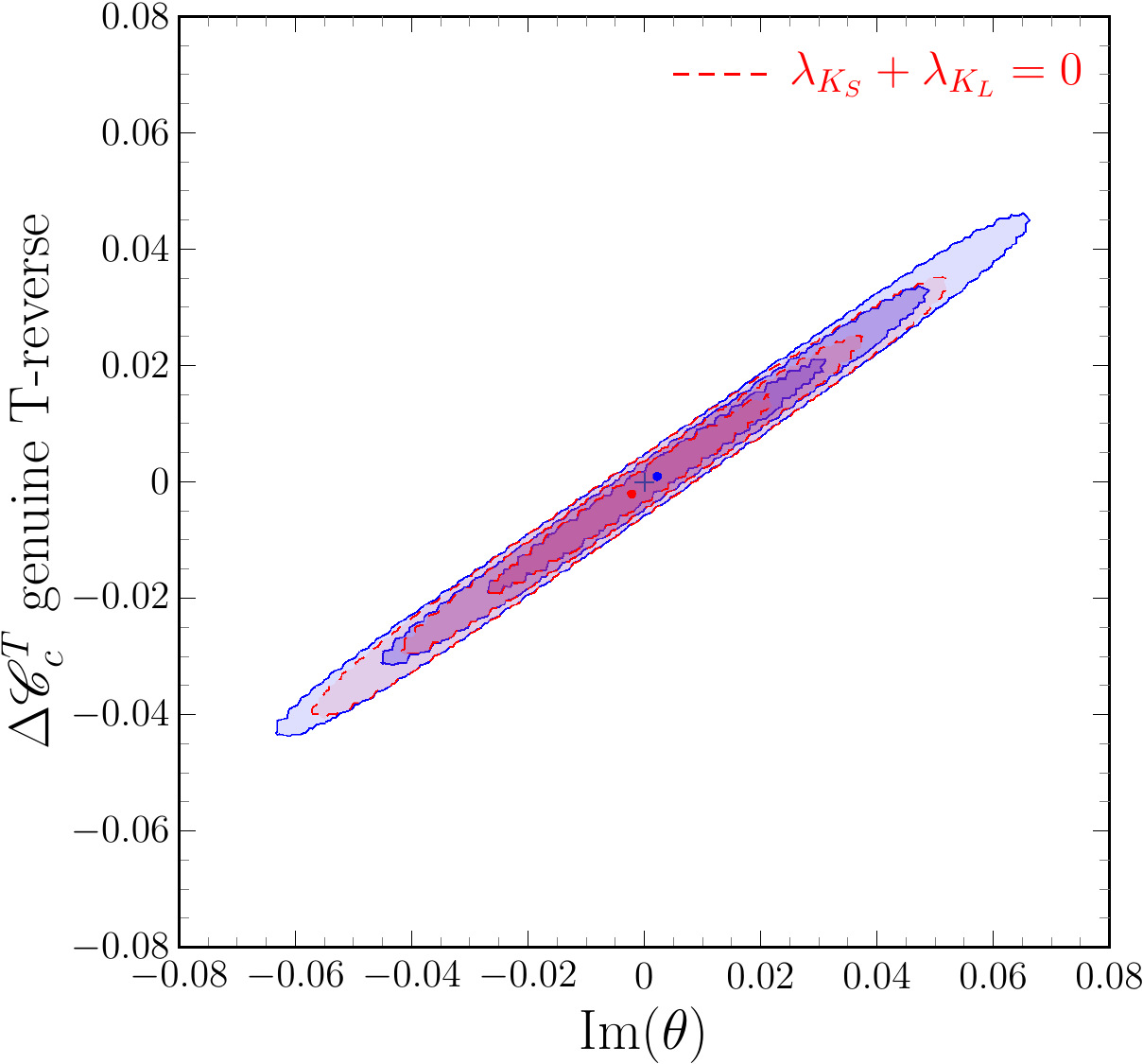}}\\
\subfigure[$\DCcCPT$.]{\includegraphics[width=0.23\textwidth]{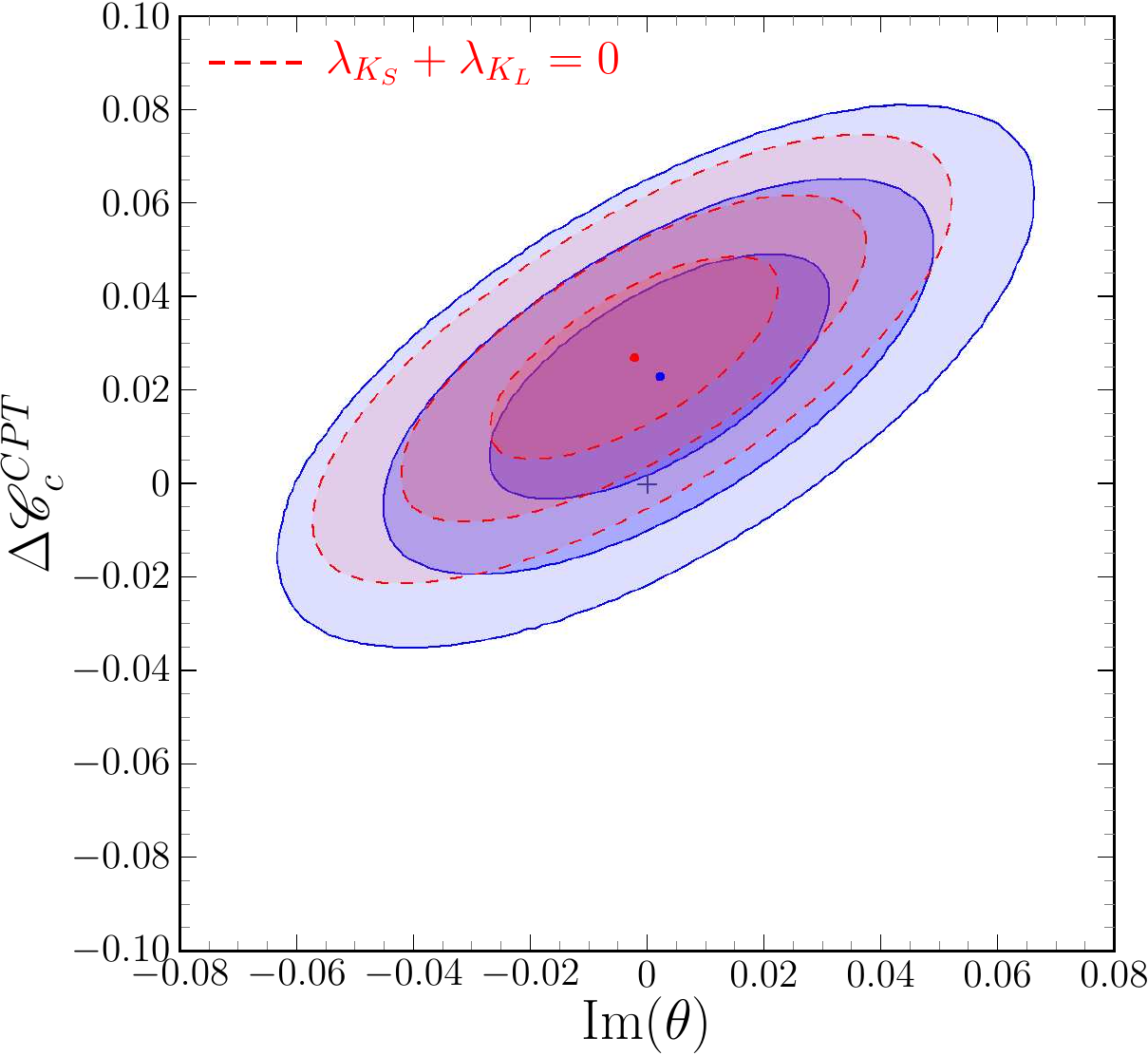}}\ 
\subfigure[Gen. T-rev. $\DCcCPT$.]{\includegraphics[width=0.23\textwidth]{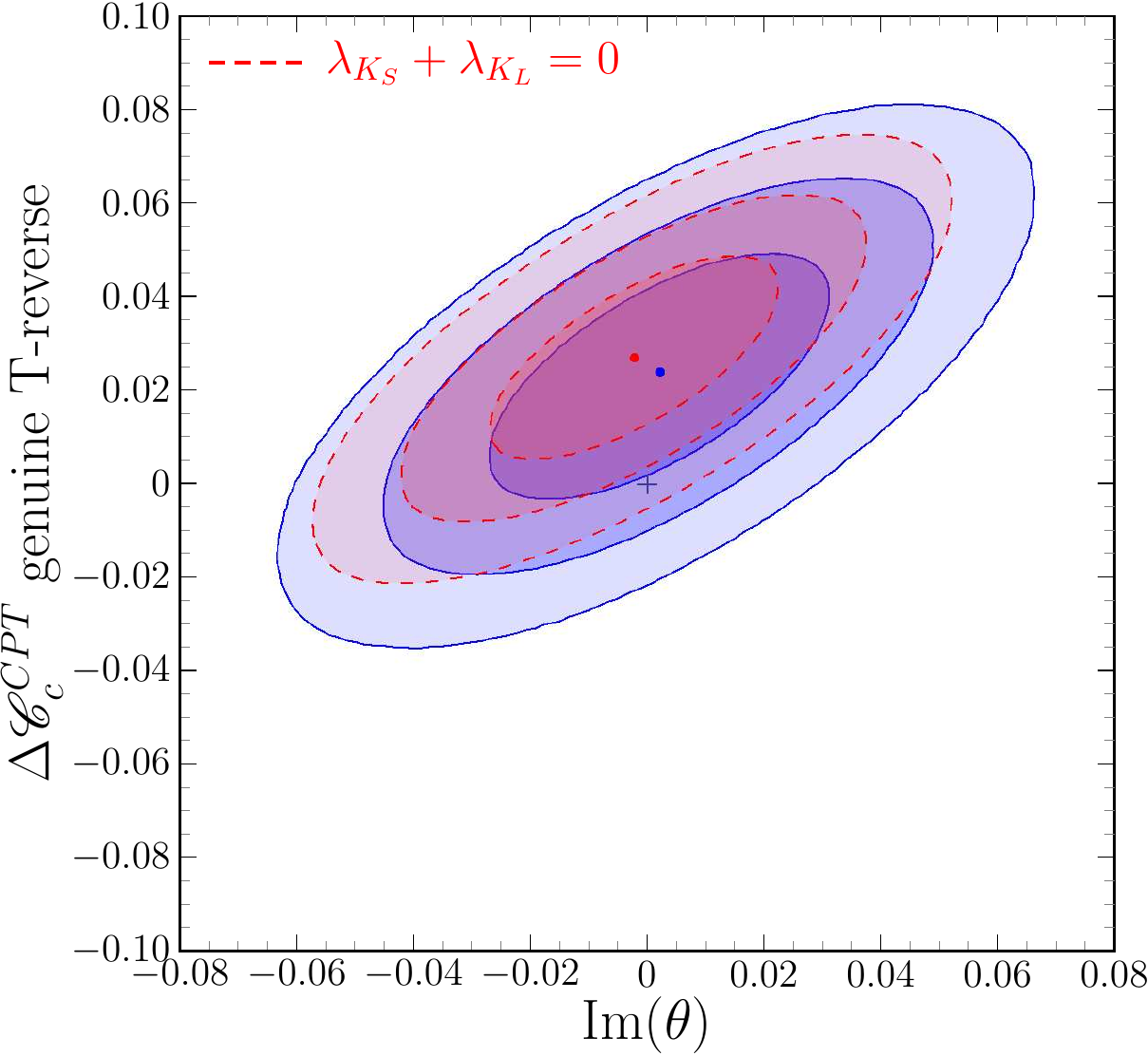}}\ 
\subfigure[$\DScCPT$.]{\includegraphics[width=0.23\textwidth]{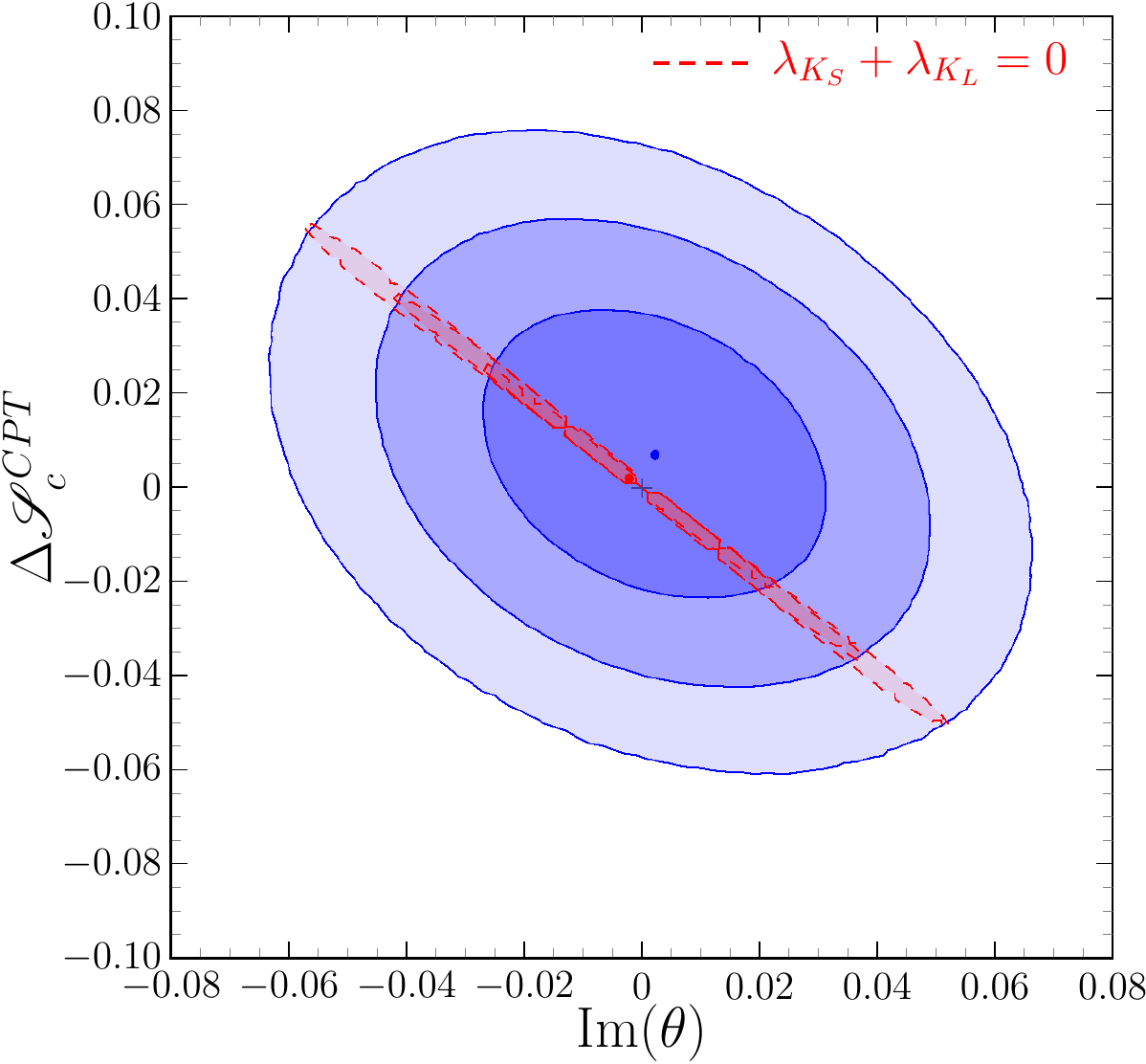}}\ 
\subfigure[Gen. T-rev. $\DScCPT$.]{\includegraphics[width=0.23\textwidth]{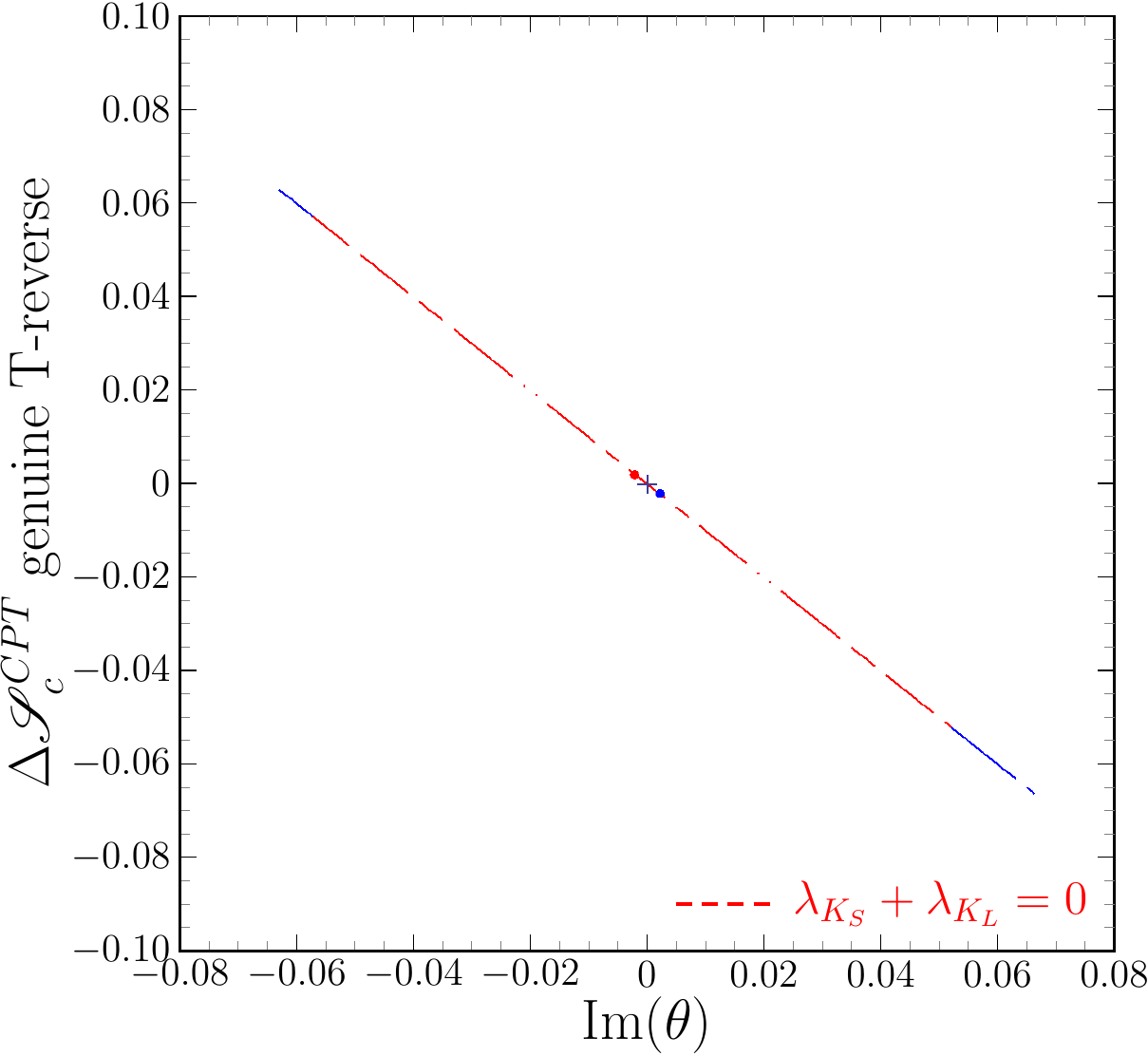}}\\
\caption{Correlations with $\im{\theta}$.}\label{FIG:ImThetaCorr}
\end{center}
\end{figure}

As shown in \refEQ{eq:theta:fits:01}, the present analysis improves on the uncertainty on $\re{\theta}$ quoted by the PDG; $\theta$, introduced in \refEQ{eq:H:params:01}, is both CP and CPT violating. It is important to stress that $\theta$ can appear not only in CPT asymmetries, but also in T and CP asymmetries, together with, respectively, CP invariant and T invariant terms. It is then interesting to explore which observables could be sensitive to $\theta$ from the theoretical point of view, and how could that translate into interesting correlations among observables and $\theta$. For $\re{\theta}$, we focus on the genuine asymmetry parameters $\DeltaCh{}$ and $\DeltaCc{}$ in equations \eqref{eq:GenuineAsym:Ch:T:01} to \eqref{eq:GenuineAsym:Cc:CPT:01} since, at leading order in $\theta$, all $\DeltaSc{}$ are insensitive to $\re{\theta}$. Attending to \refEQ{eq:GenuineAsym:Ch:CP:01}, with $\delta\simeq -5\times 10^{-4}$, $\DChCP$ could be dominated by the $-\re{\theta}\R{\KS}$ contribution; for $\DCcCP$ the situation is less clear because of the competing $\C{\KS}$ terms in \refEQ{eq:GenuineAsym:Cc:CP:01}. For $\DChT$ and $\DChCPT$, it is interesting that the $\theta$ independent terms in \refEQS{eq:GenuineAsym:Ch:T:01} and \eqref{eq:GenuineAsym:Ch:CPT:01} are suppressed by $\delta$; furthermore, since $\re{\theta}$ enters $\DChT$ with a factor $\R{\KS}+\R{\KL}$, the genuine T-reverse $\DChT$ will be interesting for $\im{\theta}$, while the genuine T-reverse $\DChCPT$ is proportional to $-\re{\theta}\R{\KS}+\im{\theta}\S{\KS}$. Similar comments apply to genuine T-reverse $\DCcT$ and $\DCcCPT$.
For $\im{\theta}$, in addition to the previous comment concerning $\DChT$, the genuine T-reverse $\DScCPT$ is, to a very good approximation, $\DScCPT\simeq -\im{\theta}$ following \refEQ{eq:GenuineAsym:Sc:CPT:01}. Notice that. although $\DScCP$ has a clean dependence in $\im{\theta}$. the dominant term $\S{\KS}\simeq -0.7$ masks this potential sensitivity\footnote{For the \BBsmix\ system, similar comments apply for decay channels $f$ with $|\Cf|\ll 1$, with, in addition, potential sensitivity to CPT Violation through $\DScCP$ if $|\Sf|\ll 1$.}. In some cases, the correlations persist partially even in the presence of fake contributions to T and CPT asymmetries. Figures \ref{FIG:ReThetaCorr} and \ref{FIG:ImThetaCorr} illustrate and confirms the previous discussion. It is to be said that, on top of the theoretical expectations, the actual experimental input is central to shape the sensitivity to $\theta$, including in particular the fact that the decay channels including $\KL$ give larger uncertainties than their counterpart with $\KS$, and thus CP asymmetries with the $\KS$ could be, a priori, better suited to uncover the presence of $\theta$.

\clearpage
\section{Conclusions\label{SEC:Conc}}

A separate direct evidence for T, CP, CPT symmetry violation needs the precise identification of \emph{genuine} Asymmetry Parameters in the time evolution of intensities between the two decays in a B factory of entangled neutral $\Bd$-meson states. By \emph{genuine} we mean a set of observables, for each symmetry, in yes-no biunivocal correspondence with symmetry violation. In this paper such a goal has been accomplished, and their values have been obtained from the BaBar measurements of the Flavour-CP eigenstate decay channels. 

In the course of this study several important results are worth mentioning, including both genuine plus possible fake effects:
\begin{itemize}
\item The meson states $B_\pm$ filtered by the observation of the CP eigenstate decay channels $J/\Psi K_\mp$ are indeed orthogonal, with extracted values for non-orthogonality $\epsilon_\rho =-0.023\pm 0.013 $, $\epsilon_\beta=0.013\pm 0.040$. $B_\pm$ are to be used as the meson states, together with $\Bd$, $\Bdb$, to obtain the transition probabilities for the asymmetry parameters.
\item The condition allowing to use Motion Reversal Asymmetry as genuine Time Reversal Asymmetry, not only with the exchange of initial and final meson states but using T-transformed states, is well satisfied with a resulting value $\rho = 1.021\pm 0.032$ in \refEQ{eq:Lambdas:Param:01}. 
Similarly for CPT Reversal Asymmetry. In addition, there is consistency for no wrong flavour in the decays, as required by \refEQ{eq:NWS:02}.
\item With any normalization in the time dependence of the intensities, a non-vanishing Asymmetry Parameter between the symmetry transformed transition probabilities is a proof of symmetry violation. However, a yes-no biunivocal correspondence is only valid with the precise connection between transition probabilities between meson states and experimental double decay rate intensities of Table \ref{TAB:Transitions} (see equation \eqref{eq:Filtering:ReducedIntensity:01}). The results obtained for these genuine Asymmetry Parameters are shown in Table \ref{TAB:fit:FULL}. 
The two extracted values for $\DScT=-0.687\pm 0.020$ and $\DScCP=-0.680\pm 0.021$ are independent direct demonstrations, with high statistical significance, of T violation and CP violation respectively. These values are compatible with the SM expectations $\DScT=\DScCP=-\sin(2\beta)=-0.682\pm 0.018$.
\item The information content of the nine genuine Asymmetry Parameters, three terms in the time dependence for each of the three T, CP, CPT symmetries, is different and we invite the reader to scrutinize the right hand sides of \refEQS{eq:GenuineAsym:Ch:T:01}-\eqref{eq:GenuineAsym:Sc:CPT:01} to identify the precise combinations of the WWA-parameters to these observables. In particular, it is crucial that the non-vanishing Asymmetry Parameters $\DScT$ and $\DScCP$ are a priori independent. In Figure \ref{FIG:Babar:Ttrue-CP} one can see that they are indeed different.
The independent information of CP violating and T violating asymmetry parameters is even more apparent in figure \ref{FIG:Babar:Ttrue-CP:Cc} where $\DCcCP$ is plotted vs. $\DCcT$.
\item From our analysis there is a 2$\sigma$ effect for the CPT-violating WWA parameter $\re{\theta}$, leading to a diagonal mass-term difference between $\Bd$ and $\Bdb$ given in \refEQ{eq:theta:fits:02}. This indication is illustrated in Figure \ref{FIG:ReImTheta}. Even with it, if the result is interpreted as an upper limit for CPT-violation, it is the best one for the $\Bd$-system and obtained from the same Flavour-CP eigenstate double decay products which have demonstrated direct evidence for CP and for T violation.
\item Last but not least, we also have an important implication for facilities without entangled states, like the LHCb experiment at LHC with high statistics. The three genuine Asymmetry Parameters for CP violation can be addressed in this experiment. 
The term $\DCcCP\,\cos(\Delta M\,t)$, even in the time dependence, may be dominated by a CPT violating contribution in this set of transitions, so we propose the disentanglement of this most interesting term at LHCb. 
\end{itemize}
Whereas all the precision measurements discussed in this paper for genuine Time Reversal violation and CPT violation Asymmetry Parameters in Flavour $\rightleftarrows$ CP eigenstate transitions need entanglement, as in the upgraded Belle experiment, the CP violation Asymmetry Parameters do not. We have insisted on the interest of separating out the different even and odd terms in the time dependence at LHCb for these transitions in the $B_d$ system.






 \section*{Acknowledgments\label{SEC:Ack}}
The authors acknowledge financial support from the Spanish MINECO 
 through Grants FPA2015-68318-R, FPA2014-54459-P and the \emph{Severo Ochoa} Excellence Center Project SEV-2014-0398, and from \emph{Generalitat Valenciana} through Grants PROMETEOII/2013/017 and PROMETEOII/2014/049.



%

\clearpage
\providecommand{\href}[2]{#2}\begingroup\raggedright\endgroup






\end{document}